\documentclass[12, preprint]{aastex}

\usepackage{natbib}
\usepackage{float}
\usepackage{url}
\newcommand{\Av}{A$_{\rm V}$\,}
\newcommand{\Td}{T$_{\rm d}$\,}
\newcommand{\kn}{\,$\kappa_\nu$\,}
\newcommand{\kv}{\,$\kappa_V$\,}
\providecommand{\e}[1]{\ensuremath{\times 10^{#1}}}
\newcommand{\thco}{$^{13}$CO\,}

\newcommand{\rfrac}[2]{{}^{#1}\!/_{#2}}
\newcommand{\cm}{\rm\, cm}

\newcommand{\erg}{{\rm\, erg}}
\newcommand{\km}{{\rm\, km}}
\newcommand{\ps}{\mbox{$\s^{-1}\,$}}
\newcommand{\pcc}{\mbox{$\cm^{-3}\,$}}
\newcommand{\s}{{\rm\, s}}
\newcommand{\sr}{{\rm\, sr}}
\newcommand{\psr}{\mbox{$\sr^{-2}\,$}}
\newcommand{\g}{{\rm\thinspace g}}
\newcommand{\scmpg}{\mbox{$\cm^{2} \g^{-1}\,$}}
\newcommand{\brightsr}{\mbox{\erg \pscm \psr \ps}}
\newcommand{\kmps}{\mbox{$\km\ps\,$}}
\newcommand{\pscm}{\mbox{$\cm^{-2}\,$}}

\newcommand{\pccm}{\mbox{$\cm^{-3}\,$}}
\newcommand{\scm}{\mbox{$\cm^{2}\,$}}
\newcommand{\water}{\mbox{{\rm H}$_2${\rm O\,}}}
\newcommand{\htwo}{\mbox{{\rm H}$_2$\,}}
\newcommand{\ammonia}{\mbox{{\rm NH}$_3$\,}}

\shorttitle{Herschel dust emission from starless cores}
\shortauthors{Wagle et al.}

\begin{document}

\title{Herschel dust emission as a probe of starless cores mass: \\
		MCLD 123.5+24.9 of the Polaris Flare}

\author{
G. ~A. Wagle\altaffilmark{1}, 
Thomas ~H. Troland\altaffilmark{1}, 
Gary ~J. Ferland\altaffilmark{1, 2}, and 
Nicholas P. Abel\altaffilmark{3}}

\email{gururaj.wagle@uky.edu}

\altaffiltext{1}{University of Kentucky, Lexington, KY 40506, USA}
\altaffiltext{2}{The Queen’s University of Belfast, Belfast, BT7 1NN, UK}
\altaffiltext{3}{University of Cincinnati, Clermont College, Batavia, OH, 45103, USA}

\begin{abstract}
We present newly processed archival Herschel images of molecular cloud MCLD 123.5+24.9 in the Polaris Flare. This cloud contains five starless cores. Using the spectral synthesis code Cloudy, we explore uncertainties in the derivation of column densities, hence, masses of molecular cores from Herschel data. We first consider several detailed grain models that predict far-IR grain opacities. Opacities predicted by the models differ by more than a factor of two, leading to uncertainties in derived column densities by the same factor. Then we consider uncertainties associated with the modified blackbody fitting process used by observers to estimate column densities. For high column density clouds (N(H) $\gg$ 1\e{22} \pscm), this fitting technique can underestimate column densities by about a factor of three. Finally, we consider the virial stability of the five starless cores in MCLD 123.5+24.9. All of these cores appear to have strongly sub-virial masses, assuming, as we argue, that \thco\ line data provide reliable estimates of velocity dispersions. Evidently, they are not self-gravitating, so it is no surprise that they are starless.
\end{abstract}


\section{Introduction}
Column densities, hence, masses of molecular clouds are estimated from observations of trace constituents of the clouds. Commonly used tracers include far-infrared (FIR) dust emission, millimeter wavelength CO line emission, and optical and near-infrared dust extinction. Dust extinction measurements are limited in sensitivity and spatial resolution by the availability of background stars, especially at high latitudes. CO line emission depends upon chemical networks, reaction rates, desorption, and adsorption (freeze-out) rates onto grain surfaces, molecular excitation and optical depth effects. In comparison, FIR dust emission is a rather straightforward thermal process depending upon the dust opacity and temperature. Observers commonly derive N(H) for molecular clouds by fitting FIR brightnesses to a modified blackbody function for which key parameters include the gas-to-dust ratio, the dust temperature \Td  (assumed constant along the line-of-sight), and the dust opacity function\kn \citep[e.g.][hereafter, WT10]{WardTh10}. Performed on a pixel-by-pixel basis, this process leads to images of N(H) and \Td  derived from images of FIR brightness in several wavelength bands. 

\defcitealias{WardTh10}{WT10}

Despite its relative simplicity, the FIR fitting technique is susceptible to several uncertainties. Among them are uncertainties in the gas-to-dust ratio, and in the dust opacity function. The technique may also be affected by variations in \Td along the line of sight, variations that lead to large variations in dust emissivity. A variation in the estimation of the line of sight averaged dust temperature by $\pm$2 K leaves the column density, and hence, the mass uncertain to a factor of two \citep{Launhardt13}.  Various recent attempts have been made in the literature to understand the effects of these uncertainties \citep[][etc.]{Schnee06, Shetty09, Kelly12, Ven13}. Many of these authors, however, do not account for the true nature of variation of dust temperatures. Most observers attempt to fit the SEDs to observations via variations in the spectral index $\beta$ of the opacity function \& a single dust temperature. \citeauthor{Shetty09} acknowledge that the estimated dust temperature through various such methods is only representative, and provide an upper limit for the coldest temperature along the line of sight. Here we use the spectral synthesis code Cloudy \citep{c13ref} to explore uncertainties in the values of N(H) that are derived from observations of FIR brightnesses. Specifically, we consider $(a)$ uncertainties in \kn and $(b)$ the effects of declining \Td into a starless core that is externally heated by the interstellar radiation field (ISRF). These uncertainties are important to quantify because uncertainties in N(H) for starless cores translate into uncertainties in their masses. Uncertainties in masses, in turn, can create uncertainties in our understanding of the virial stabilities of the cores, hence, in their future evolution. In this study, we use starless cores in the Polaris Flare as examples, and we consider their virial status based upon estimates of N(H). However, our results apply more generally to starless cores elsewhere in the Galaxy. In a future study, we will use Cloudy to explore uncertainties in the relationship between observed molecular emission and N(H) in starless cores.

The Polaris Flare is a translucent molecular cloud situated at Galactic latitude $\sim$25$\degr$ that was discovered by \citet*{Heith90}. These authors put an upper limit on the distance of 240 pc. Thereafter, most authors have adopted a distance of 150 pc, which we assume here. At this distance, the cloud lies within the Galactic molecular disk. MCLD 123.5+24.9 (hereafter, MC123) is one of the denser regions within the Polaris Flare. MC123 is well observed in molecular tracers and is gravitationally unbound \citep{Bensch03, HBF, Heith02, Heith08, Shim12}. In addition, recent observations of the FIR dust emission from MC123 by the Herschel Space Telescope \citep[see][and \citetalias{WardTh10}]{Andre10} identified several molecular cores. No IRAS or Spitzer point sources are associated with any of the MC123 cores \citetalias{WardTh10}. Therefore, these cores are starless \citep{Heith08}.

The structure of this paper is as follows. In \S2, we present our new reductions of archival Herschel data for MC123. The nature of this region, including its five starless cores, is outlined. We use these data to study the properties of the MC123 starless cores.  Later in the manuscript, we use the data to explore uncertainties in determination of column densities of starless cores in general. In \S3, we describe the basic features of Cloudy models of starless cores that are heated externally by the ISRF. In \S4, we present several dust grain models from which dust opacities have been calculated by Cloudy. In \S5 we use Cloudy to explore the uncertainties in N(H) derived from FIR observations via the standard observers’ fitting process. Section \S6 outlines our conclusions regarding the virial stability of the starless cores of MC123. Finally, \S7 summarizes our conclusions.

\section{Archival Herschel Observations of the Polaris Flare}

The molecular cloud MC123 of the Polaris Flare is an elongated structure of size $\approx$ 1.5$\times$0.5 pc with an average visual extinction, AV $\approx$ 0.5\,--\,0.8 mag \citep{Bensch03, HBF, Shim12}. Therefore, N(H) $\approx$ 1\,--\,2\e{21}\pscm, assuming a standard ratio of N(H)/AV = 1.8\e{21}\pscm mag$^{-1}$. MC123 shows strong extended IRAS 100 $\micron$ emission and is a local maximum in the $^{12}$CO(1$\to$0) line intensity map of the cloud. MC123 was observed with the Herschel Space Telescope as a part of the Gould Belt Survey \citep{Andre10}. These science demonstration phase (SDP) observations were performed at 70 $\micron$ and 160 $\micron$ with Photodetector Array Camera and Spectrometer \citep[PACS;][]{PACSref}, and 250 $\micron$, 350 $\micron$ and 500 $\micron$ with Spectral and Photometric Imaging Receiver \citep[SPIRE;][]{SPIREref}. From these data, \citetalias{WardTh10} published 24$\arcsec$ resolution images of MC 123 at 160 $\micron$, 250 $\micron$, and 350 $\micron$. The morphology of the images is very similar to that observed in \thco by \citet*{HBF}. \citetalias{WardTh10} identified five core regions within MC123 and numbered them in order of increasing Galactic longitude. These cores are molecular and dense. \citetalias{WardTh10} estimated the mean molecular hydrogen density, n(H$_{\rm 2}$) $\sim$10$^{5}$\pccm. They also derived a dust temperature, \Td $\sim$10 K and peak-hydrogen column density, N(H$_{\rm 2}$)$_{peak} \sim$10$^{22}$\pscm for the cores.

For this study, we combined the SDP data described above for MC123 with data from the Guaranteed Time Key Project (KPGT) for the same field, both available at Herschel Science Archive (HSA). For a given Herschel band, the SDP and the KPGT data sets each contain two observations of the MC123 field with cross-linked scans. Therefore, we combined these four observations in a given band into a single image.

To process the SPIRE data (250 $\micron$, 350 $\micron$, 500 $\micron$), we used the latest version of the Herschel Interactive Processing Environment \citep[HIPE, v11.0.2,][]{HIPEref}. Processing with the new version ensured that the images are calibrated using the latest calibration tree (v11.0) which has improved since WT10. We used a plug-in to HIPE, called SPIRE Photometer Interactive Analysis \citep[SPIA, v1.11.1,][]{SPIAref}. For each Herschel band, the data were destriped and the extended emission images were produced. These images were zero-point corrected to take into account the absolute offset on SPIRE images of the radiative contribution from the telescope mirror, based on the cross-calibration with HFI 545 and 857 $\micron$ images from Planck mission. This is a linear offset applied to the entire map. The maps were calibrated to surface brightness units (MJy/Sr) by dividing by the effective beam solid angle for a constant $\nu$S$_\nu$ source. These maps were then corrected for the variation of the relative spectral response function (RSRF) and aperture efficiency, and for the effective beam solid angle for the true spectrum of the source, by applying photometric color-correction parameters to account for the difference between the observed spectrum and the calibrated (flat) spectrum. This correction is within $\pm$2\% for the three bands. The contribution from the cosmic microwave background (CMB) dipoles is removed during the processing (private communication with Herschel Science Center help desk). The absolute calibration uncertainty in the SPIRE maps due to the use of Neptune model is $\pm$5.5\%. In addition, there is an uncertainty of $\pm$4\% due to uncertainty in the measured beam area. The total calibration uncertainty of the SPIRE maps is better than 15\%. \citep[See][for further details]{SPIREhb}.

\begin{deluxetable}{ccccc}
\tablecolumns{5}
\tablecaption{\label{tab1} Herschel Space Telescope data}
\tablewidth{0pt}
\tabletypesize{\small}
\tablehead{
	\colhead{$\lambda$ ($\micron$)} & \colhead{Instrument} & 
		\colhead{Reduction Software} & \colhead{Pixel Size ($\arcsec$)} & \colhead{FWHM ($\arcsec$)}\\
	\colhead{(a)} & \colhead{(b)} & \colhead{(c)} & \colhead{(d)} & \colhead{(e)}}
\startdata
160 & PACS & UNIMAP & 4.5 & 13\\
250 & SPIRE & HIPE & 6 & 18\\
350 & SPIRE & HIPE & 10 & 24\\
500 & SPIRE & HIPE & 14 & 35\\
\enddata
\end{deluxetable}

PACS data (70 $\micron$, 100 $\micron$, 160 $\micron$), were processed with the external Unimap software \citep{Piazzo12, Piazzo15a, Piazzo15b}. Unimap produces high quality images implementing a full pipeline, starting from the level 1 data of the standard pipeline. We use the GLS maps produced by the software. The maps were calibrated in Unimap to surface brightness units (MJy sr$^{-1}$). The offset for the maps produced by Unimap is arbitrary. We subtracted the contribution from a relatively “empty” region in the field of the image. This also ensures that the CMB contribution is subtracted from the brightness value we use. It is worth noting, however, that the theoretical value for CMB contribution at PACS wavelengths is more than a thousand times smaller than the modified black body dust emission at the dust temperatures considered here. The PACS 160 $\micron$ data yielded an useful image. However, the signal-to-noise ratio in the shorter wavelength PACS bands (70 and 100 $\micron$) was too low to be useful. This outcome is not surprising since dust emission at these wavelengths should be negligible in cold molecular clouds. It might also be advisable to exclude fluxes shortward of 100 $\micron$ while deriving dust properties of molecular cores, for various reasons \citep[see][]{Shetty09}. Therefore, we make no further reference to the 70 and 100 $\micron$ PACS bands. Unimap (like all other Herschel mappers, including Scanamorphos) is currently not able to produce an accurate estimate of the uncertainties (Private communication with L. Piazzo). \citet{Paladini12} found that the PACS maps at 160 $\micron$ agree to within $\sim$5-20\% with MIPS 160 $\micron$ maps, which have similar angular resolution and a well-documented calibration accuracy. We put the uncertainty in PACS maps at 20\% to be on the safe side.

\begin{figure}
\includegraphics[scale=0.58]{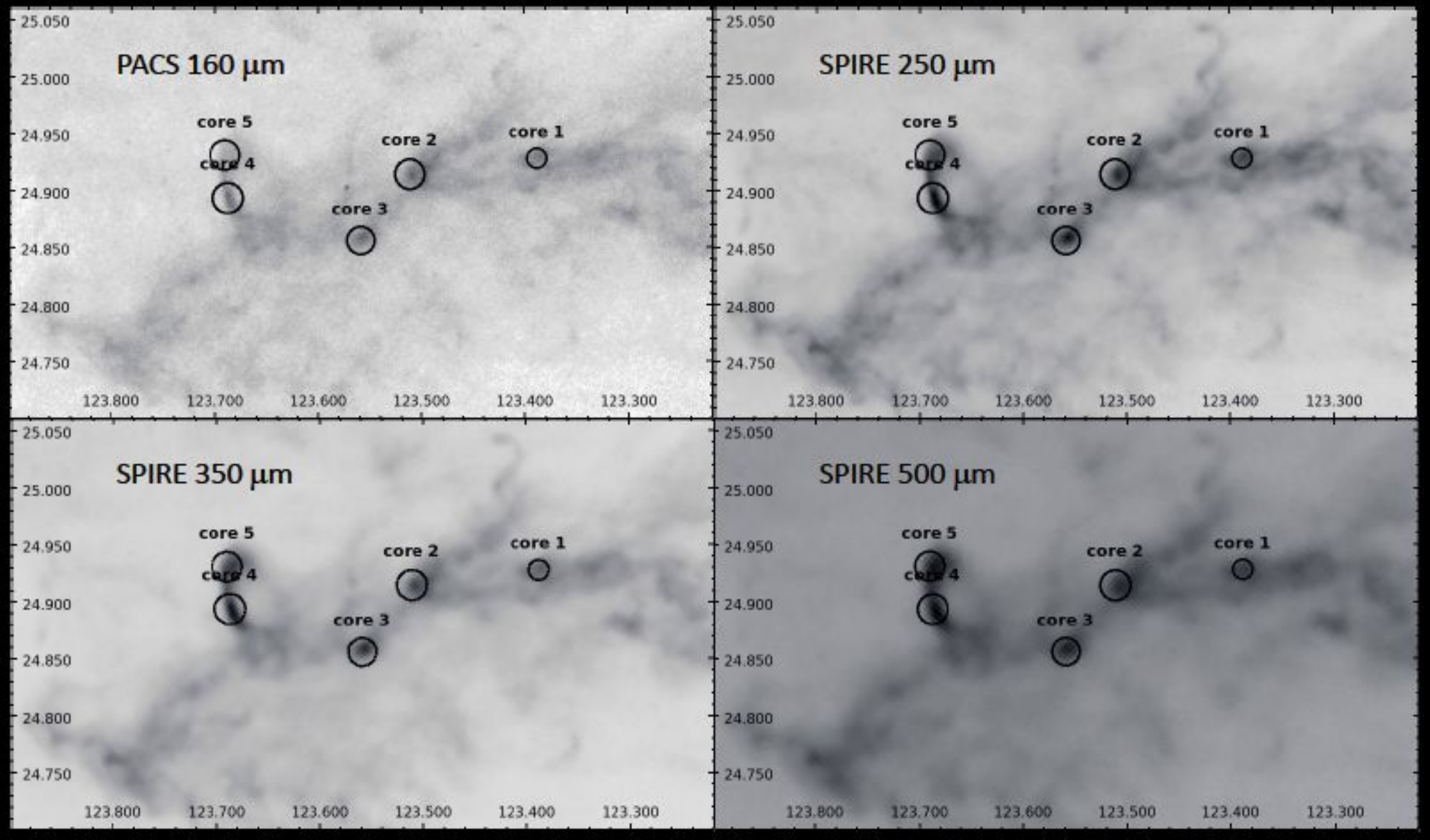}
\caption{
Processed dust emission maps at the four longest wavelength bands. Upper row: 160 $\micron$ PACS, and 250 $\micron$ 
SPIRE. Lower row: 350 $\micron$ and 500 $\micron$ SPIRE. The circles represent the cores identified by \citeauthor{WardTh10} 
Spatial resolutions of these unsmoothed images are given in Table 1.}
\label{fig:maps}
\end{figure}

Basic parameters of the Herschel images are listed in Table \ref{tab1}. Figure \ref{fig:maps} shows the processed images for the four Herschel bands that yielded useful results. The images for the 160  $\micron$, 250  $\micron$ and 350  $\micron$ bands are similar to those shown by \citetalias{WardTh10}. Figure \ref{fig:maps} also includes an image of the 500  $\micron$ band. In Table \ref{tab2}, we list surface brightnesses (MJy sr$^{-1}$) at the central pixel of each core identified by citetalias{WardTh10} and for each Herschel band. Note here that these surface brightnesses have contribution from the inter-core region included in them, as our models do, too (see section 3.1). For the purposes of extracting brightness information in Table \ref{tab2}, we smoothed the 160  $\micron$, 250  $\micron$ and 350  $\micron$ images of Figure \ref{fig:maps} to the 35$\arcsec$ resolution of the 500  $\micron$ map, using a subroutine imsmooth in a standalone software WCSTools \citep[v3.8.7,][]{WCSref}. However, the images in Figure \ref{fig:maps} are unsmoothed, having the resolutions listed in  Table \ref{tab1}.

\begin{figure}
\includegraphics[scale=0.68]{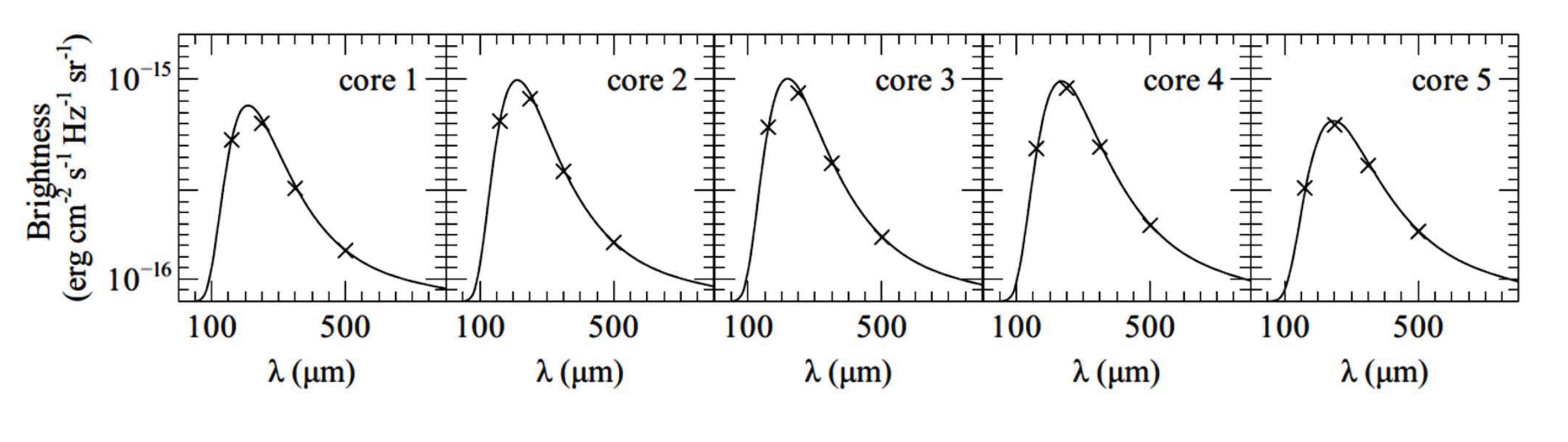}
\caption{
The SED fits to dust emission observations as explained in the text are shown here. The crosses show the peak surface brightness measured in each band for each core}
\label{fig:fits}
\end{figure}

We have fitted the brightnesses listed in Table \ref{tab2} to the modified blackbody function used by WT10 and other observers. (\citetalias[See][]{WardTh10} equations 1 and 2.) This technique returns values of N(H) and the dust temperature \Td for each MC123 core, listed in Table \ref{tab2}. \Td is assumed constant along the line-of-sight, an assumption that is not likely to be correct. Nonetheless, the modified blackbody function fits the four observed FIR brightnesses for each core very well. (See \S5.) Figure \ref{fig:fits} shows these fits for all of the cores. We note a small conceptual difference between our fits and those of \citetalias{WardTh10}. \citetalias{WardTh10} fitted flux densities (in Jy) for each of the MC123 cores. As described in the notes to their table 1, these flux densities are brightnesses integrated out to the contours of half peak brightness. Our fits, however, are to the peak brightnesses (in MJy sr$^{-1}$) of each core, extracted from images smoothed to a common 35$\arcsec$ resolution (0.03 pc at 150 pc, see our Table \ref{tab2}). Since the FWHM dimensions listed in \citetalias{WardTh10} table 1 are all close to the 0.03 pc spatial resolution of our smoothed images, their approach of fitting to flux densities should be closely equivalent to our approach of fitting to peak brightnesses.

\begin{deluxetable}{rccccc}
\tablecolumns{6}
\tablecaption{\label{tab2} The Peak Brightness in Dust Emission for each core}
\tablewidth{0pt}
\tabletypesize{\small}
\tablehead{
	\colhead{Parameters} & \colhead{Core1} & 
		\colhead{Core 2} & \colhead{Core 3} & \colhead{Core 4} & \colhead{Core 5}\\
	\colhead{(a)} & \colhead{(b)} & \colhead{(c)} & \colhead{(d)} & \colhead{(e)} & \colhead{(f)}}
\startdata
Galactic Longitude, l & 123.388 & 123.511 & 123.559 & 123.687 & 123.690\\
Galactic Latitude, b & +24.928 & +24.915 & +24.856 & +24.894 & +24.931\\
RA, $\alpha$ (J2000) & 1$^h$34$^m$01.9 & 1$^h$44$^m$51.6 & 1$^h$47$^m$40.8 & 1$^h$59$^m$42.7 & 2$^h$00$^m$58.7\\
Declination, $\delta$ (J2000) & $+$87$\degr$45$\arcmin$42$\arcsec$ & $+$87$\degr$43$\arcmin$35$\arcsec$ & 
	$+$87$\degr$39$\arcmin$33$\arcsec$ & $+$87$\degr$39$\arcmin$53$\arcsec$ & $+$87$\degr$41$\arcmin$58$\arcsec$\\
HWHM (pc) & 0.025 & 0.036 & 0.044 & 0.036 & 0.030\\
Distance (pc) & 150 & 150 & 150 & 150 & 150\\
B$_{\nu, \rm peak}$ (160 $\micron$) (MJy sr$^{-1}$) & 71.7 & 82.2 & 76.0 & 74.6 & 47.1\\
B$_{\nu, \rm peak}$ (250 $\micron$) & 86.1 & 95.5 & 101.9 & 109.3 & 83.8\\
B$_{\nu, \rm peak}$ (350 $\micron$) & 51.2 & 55.1 & 62.4 & 73.3 & 59.4\\
B$_{\nu, \rm peak}$ (500 $\micron$0 & 21.9 & 24.2 & 25.2 & 29.8 & 29.2\\
\Td (K) & 14 & 14 & 13 & 12 & 13\\
N(H)$_{\rm peak}$ (\e{21} \pscm) & 6 & 7 & 8 & 12 & 13 \\
Mass (M$_{\sun}$) & 0.13 & 0.33 & 0.58 & 0.50 & 0.40\\
\enddata
\tablecomments{ 
The surface brightness for each core at 160 $\micron$ PACS band, and 250, 350, and 500 $\micron$ SPIRE bands are measured at the position of the brightest pixel in the 500 $\micron$ image; Galactic co-ordinates are listed in this table. Surface brightnesses at 160, 250 and 350 $\micron$ come from images that have been smoothed to the 35$\arcsec$ resolution of the 500 $\micron$ image. The surface brightnesses have been corrected for variations in telescope beamwidth across each band (“color correction”), a correction that amounts to less than 2\%. The uncertainty in brightness is $\pm$15\% for SPIRE and $\pm$20\% for PACS 160 $\micron$, as described in the text. The HWHM is the geometric mean half-width at half maximum measured at the peak of the core in our unsmoothed 250 $\micron$ maps. A dust temperature (\Td), peak hydrogen column density, and mass were derived as described in the text.}
\end{deluxetable}

Our fitted values for N(H) are about one half those reported by \citetalias{WardTh10} for each of the five MC123 cores. Also, our derived values for \Td are about 15\% higher than those derived by \citetalias{WardTh10}. These differences may arise from small differences in the two data sets related to details of the calibration and reduction processes. In addition, these differences may be related to differences in sampling of the data to derive flux densities vs. peak brightnesses, as described above. Since our data set includes more observations and it has been subject to a more recent reduction pipeline, we suspect that our results are more reliable. Nonetheless, the factor of two differences in N(H) derived independently from the \citetalias{WardTh10} and the present data sets suggests that systematic errors may limit the accuracy of column density measurements to of order a factor of two. We also include in Table \ref{tab2} estimates of the masses for each of the cores. We have used the simple cylindrical approximation in which M $\propto \pi$R$^2$N(H), with a scale factor of 1.4 to account for He mass. Values for the core radii R are taken from the angular radii HWHM of Table \ref{tab2}, converted to linear units at the adopted distance of 150 pc. Core masses in Table \ref{tab2} are generally quite comparable to those derived by \citetalias{WardTh10}. The differences likely reflect possible differences between the two data sets described above as well as differences in geometrical assumptions used to convert N(H) into mass.

\section{Cloudy Models of Starless Molecular Cores}
\subsection{Basic Characteristics of the Cloudy Models}
Our Cloudy models of starless cores, using the released version 13.02 \citep{c13ref}, incorporate a variety of assumptions about physical conditions and geometry. Cloudy calculates equilibrium (time independent) conditions. The models use plane-parallel (slab) geometry, with the illuminated face exposed to the interstellar radiation field (ISRF). Cloudy then calculates physical conditions in zones, beginning at the illuminated face and ending at a specified stopping value of the column density, N(H)Cloudy. From these physical conditions, Cloudy computes the brightness (MJy sr$^{-1}$) as a function of frequency of the radiation emerging from the illuminated face. Of course, molecular cores are often assumed to be spherical. However, our slab models represent a narrow column through the center of such a spherical core and parallel to the line-of-sight. The predicted brightnesses of such a model are to be compared with the peak brightnesses observed toward a core (e.g. Table \ref{tab2}), not with the integrated brightnesses across the core as a whole (i.e. the total fluxes of the core).  Our models do not predict brightness variations in the plane of the sky since our principal interest in this study is the relationship between observed FIR brightnesses and N(H). Future versions of Cloudy may incorporate spherical geometry into clouds illuminated by the external ISRF. Such models would be useful in interpreting future FIR observations of molecular cores at higher spatial resolution.

To simulate a cloud illuminated from both sides by the ISRF, we calculate all slab models up to a depth corresponding to the midpoint (i.e. N(H)Cloudy/2) of the model cloud. This calculation predicts brightnesses emerging from the front half of the slab. Then we duplicate the calculated results to account for the back half of the slab. The dust emission from the back half is attenuated by e$^{-\tau}$, where tau is the calculated continuum optical depth from the front half of the slab. However, the dust emission is optically thin, so $\tau_{\rm dust} \ll$ 1.

Other properties common to our Cloudy models include assumptions about the ISRF, the CMB, and \htwo formation and excitation. We assume that the ISRF has the SED from microwave through far-ultra violet (FUV) described by figure 2 of \citet{Black87}. We exclude H-ionizing radiation from the\citeauthor{Black87} radiation field. The integrated ISRF brightness is 2\e{-4} \brightsr or 1 Habing \citep{Habing68}. That is, G$_0$ =1. Adoption of G$_0$ =1 for the MC123 cloud is plausible given the absence of any known local enhancement in the Polaris Flare ISRF \citep{Bensch03} and the location of the Polaris Flare within the galactic molecular disk. However, we experiment with other values of G$_0$ in our models (\S5). We exclude the CMB from the models. We do so because we compare Cloudy model predictions of FIR dust emission with Herschel observations in which the CMB component has already been subtracted during the processing (\S2). In principle, exclusion of the CMB allows model temperatures to fall below 2.7 K. To preclude this possibility, we set a minimum temperature in the models of 5 K. Various theoretical and observational studies indicate that the dust temperatures at the center of the starless cores could reach a value lower than about 7 K \citep[see][and references there in]{Shetty09}. \citet{Evans01} compared the heat input due to all possible heating mechanisms to the radiative cooling for a dust grain at 5 K. They have shown that ISRF is the dominant source of heating for dust at the central regions of even an opaque core, and the gas cannot substantially raise the dust temperature. The central dust temperatures are in the range of 7-8 K in their models.  Finally, we use the large model of the H2 molecule \citep{Shaw05} although following chemical processes in the molecular cores is not the principal purpose of this study of FIR emission.

Each Cloudy model has a specified ISRF (i.e. value of G$_0$, typically ≈ 1), a specified stopping column density N(H)$_{\rm Cloudy}$ and a specified model of interstellar grains which is closely constrained by elemental abundances in the ISM (\S4). The Cloudy model then predicts emergent radiation brightness over a very broad band of wavelengths, including the FIR bands observed by Herschel. Comparisons between predicted and observed FIR emission reveal the relationships to be expected between N(H) and FIR emission, including the roles of G$_0$, the grain model and variations in \Td along the line-of-sight.

\subsection{Density Law for the Polaris Flare MC123 cores 1 and 4}

\begin{figure}
\begin{center}
\includegraphics[scale=0.8]{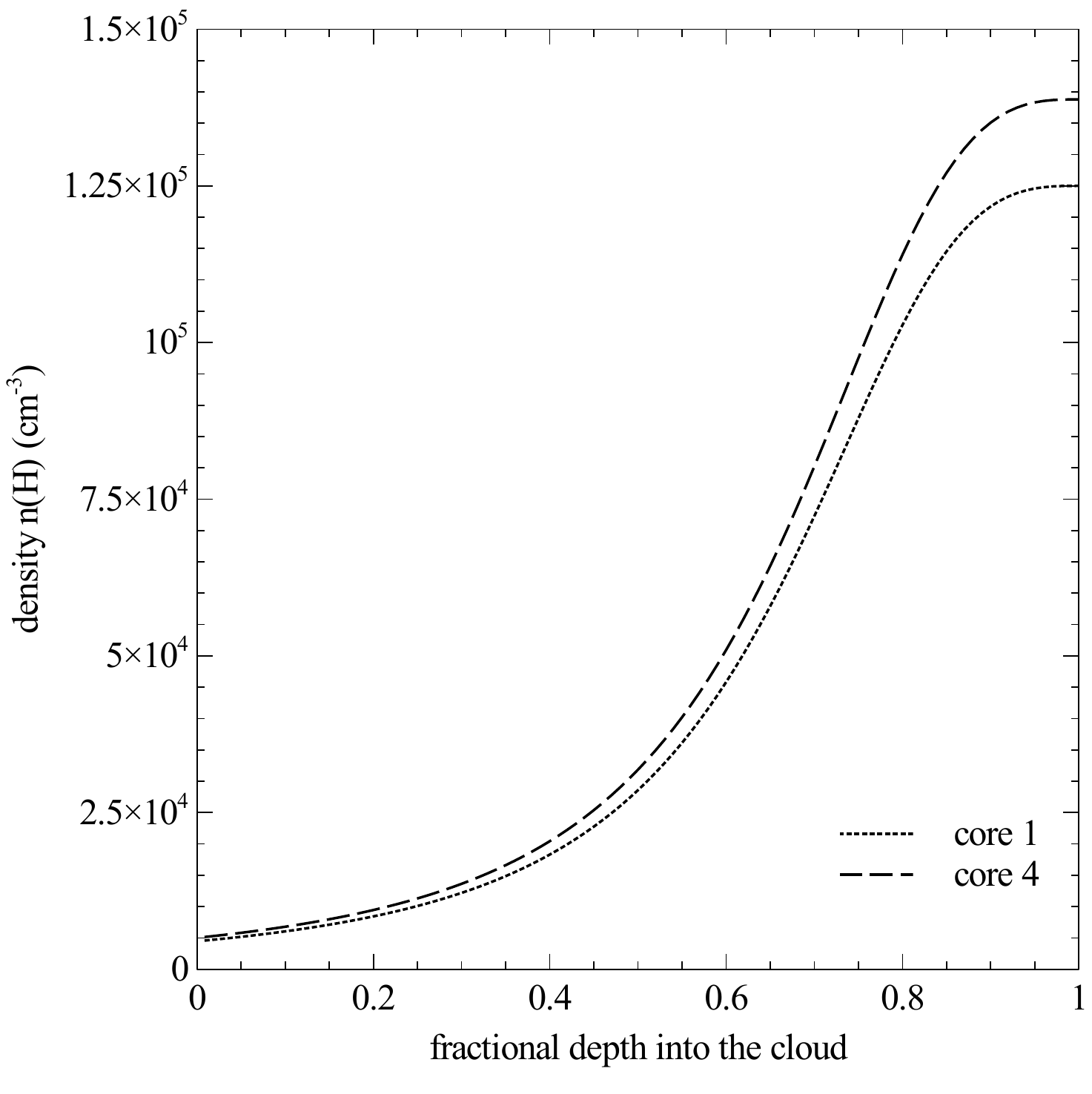}
\caption{
n(H) as a function of the fractional depth from illuminated face to the center of the core. The total model depth is two times the core radius for each core. The parameters used for the density law for each core are listed in Table \ref{tab3}.}
\label{fig:dlaw}
\end{center}
\end{figure}

\begin{deluxetable}{cccccc}
\tablecolumns{6}
\tablecaption{\label{tab3} Density Law Parameters}
\tablewidth{0pt}
\tabletypesize{\small}
\tablehead{
	\colhead{Cores} & \colhead{$\alpha$} & 
		\colhead{d$_{\rm stop}$} & \colhead{d$_{\rm scale}$} & \colhead{n$_{\rm stop}$} & \colhead{n$_{\rm scale}$}\\
	\colhead{(a)} & \colhead{(b)} & \colhead{(c)} & \colhead{(d)} & \colhead{(e)} & \colhead{(f)}}
\startdata
1 & 3 & 0.047 & 0.013 & 1.25\e{5} & 6.1\e{4}\\
4 & 3 & 0.077 & 0.025 & 1.39\e{5} & 6.8\e{4}\\
\enddata
\end{deluxetable}

Predictions of emergent FIR dust emission from model cores should not be sensitive to the run of volume density in the cores. This conclusion follows from the fact that \Td, hence, dust emissivity in a given zone of the model, depend only upon N(H) in front of the zone (i.e. between the zone and the illuminated face), not upon the density law that leads to N(H). However, with a view towards future models of starless core chemistry, and in the interest of more realistic core models, we have included density laws in our models that are designed to apply to MC123 cores 1 and 4. These two cores span the range in the sizes in MC123; core 1 is the smallest, core 4 is among the largest (Table \ref{tab2}). We used a modified version of the density law of \citet{Tafalla04}. The density, n as a function of depth, d into the cloud from the illuminated face is written as
\begin{equation}
n(d) = \frac{n_{stop}}{1+{\left(\rfrac{d_{stop}-d}{d_{scale}}\right)}^{\alpha}}
\end{equation}
where, d$_{\rm stop}$ is the depth into the cloud where the model stops the calculations, n$_{\rm stop}$ is the density at the stopping depth (center of the core), d$_{\rm scale}$ is the scale depth used to control the shape of the density law, and $\alpha$ is the scaling exponent for the density law. The value of $\alpha$ for starless cores varies between 2 and 4 \citep{Tafalla02}. We adopted a value of $\alpha$ =3 for the two cores. Density law parameters for cores 1 and 4 were chosen to match two characteristics of each core, (i) the radius to half FIR brightness (HWHM), and (ii) N(H), both listed in Table \ref{tab2}. For each of the two cores, we chose values of d$_{\rm stop}$, d$_{\rm scale}$, and n$_{\rm stop}$ so that the integrated density law reproduced the specified N(H). The resulting density law for each core has a total depth d$_{\rm stop}$ of about two times the radius (i.e. HWHM in Table \ref{tab2}). For each core, n(H) varies from a few times 10$^3$  \pcc at the illuminated face to a few times 10$^5$ \pcc at n$_{\rm stop}$. (See Table \ref{tab3} \& Figure 3.) The lower density regions described by the density law, closer to the illuminated face, well outside the specified core radii represent the “inter-core” gas in the vicinity of the cores of MC123. citet{Heith08} used multiple HC$_3$N lines to constrain the density and found (1.1 $\pm$ 0.5) \e{5} \pcc toward the brightest peak of the HC$_3$N emission in core 4. This density and the dimension over which they derived it from the observations are quite consistent with the density laws shown in Figure 3. The parameters used for the density law for the two cores are listed in Table \ref{tab3} below, along with the average density for each core.

\section{Grain Models, FIR Dust Opacity, and Column Densities}

Values of N(H) derived from FIR observations are inversely proportional to the assumed values of \kn, values that cannot be directly measured. Instead, the run of \kn in the FIR must be calculated from a grain model, which, in turn, is utilized in Cloudy models to predict emission from molecular regions. A grain model specifies the size distribution and compositions of the grains, constrained by the elemental abundances available for grain formation. A grain model typically includes grains of different types. Finally, a grain model makes use of (or calculates) refractive indices of the grains to determine \kn. Cloudy can construct multi-component grain models that include grain sizes in the approximate range 1‒1000 nm. The smaller grains dominate grain surface processes, such as formation of \htwo and other molecules. These grains affect Cloudy predictions of molecular line strengths, not directly relevant to the present study of FIR continuum emission. The larger grains contribute most of the dust mass, and they re-emit stellar FUV radiation in the FIR. These grains determine Cloudy predictions of FIR emission. We have used Cloudy to construct several detailed, multi-component grain models. For these models, Cloudy uses effective medium theory (EMT) to calculate refractive indices of mixed grain materials, and Cloudy also incorporates refractive indices from other sources (see appendix for further details). Cloudy then uses Mie theory to calculate grain opacities, \kn as a function of frequency from the refractive indices. Our procedures are similar to those of \citet[ hereafter, Pr93]{Preb93}, 
\defcitealias{Preb93}{Pr93}
and we have used refractive index data from that work. We have implemented the concept of coagulation in the form of the core mantle particles described by \citetalias{Preb93}; however, we have not considered other density, temperature or time dependent coagulation effects, or aggregate grains considered in the literature \citep[see e.g.][]{OH94}. Note that opacities for all grain models have a \kn $\propto \nu^2$ behavior in the FIR, a natural consequence of Raleigh scattering of wavelengths much longer than the grain sizes.

\begin{deluxetable}{cccc}
\tablecolumns{4}
\tablecaption{\label{tab4} Grain Types and Sizes}
\tablewidth{0pt}
\tabletypesize{\small}
\tablehead{
	\colhead{Grain Type} & \colhead{Abbreviation} & \colhead{Minimum Grain Size} & \colhead{Maximum Grain Size}\\
	\colhead{(a)} & \colhead{(b)} & \colhead{(c)} & \colhead{(d)}}
\startdata
Amorphous Carbon & aC & 0.007 & 0.03\\
Silicon Core	& Si-core & 0.04 & $\sim$1\\
\enddata
\tablecomments{
The grain sizes are in $\micron$ taken from \citet{Preb93}}
\end{deluxetable}

Observations of heavy element abundances place important constraints upon grain models as noted by \citet{Snow96} among other authors. The grain models must incorporate heavy elements, most notably, C, N, O, Mg, Si and Fe, in the proportions implied by observed depletions in the ISM and assumed cosmic abundances. The latter are usually taken to be solar abundances; we adopt those of \citet{Asplund09}. \citet{Jenkins09} compiled depletion data for various elements. He finds that lines of sight to different stars often have systematically different depletions, and he lists maximum and minimum depletions for many elements. We assume the maximum depletions reported by Jenkins for the elements. This choice is reasonable for the cold molecular gas of the Polaris Flare and other starless cores where high depletions are likely. This choice also maximizes the heavy elements available to make grains. A physically reasonable grain model should not require higher abundances of heavy elements than those implied by the maximum depletions and cosmic abundances. Such a model would overuse the elements available for grains. Likewise, a physically reasonable grain model should not require lower abundances, at least not for a cold molecular cloud where high depletions are expected. Such a model would underuse the elements available for grains; hence, it would not account for the location of elements known to be depleted from the gas. Ideally, a grain model will use all elements optimally, that is, require grain element abundances implied by cosmic abundances and depletions. Of course, cosmic abundances and observed depletions have uncertainties, leading to uncertainties in the abundances of elements available for grains. For example, cosmic abundances taken for the Sun, as adopted here, are typically 25\% higher than those derived from observations of B stars \citep[see compilation in][]{Asplund09}. In addition, depletions vary among the elements that make up grains. Mg, Si and Fe are all highly depleted in the ISM; essentially, the full cosmic abundance of each element is available for formation of silicate cores in the grain models described below. Therefore, uncertainties in grain abundances of these elements are primarily the uncertainties in cosmic abundances, of order 25\%. However, C, N and O are much less highly depleted, and N may not be depleted at all \citep{Jenkins14}. These elements are found in the dirty ice mantles of grain models described below. The grain abundances of these elements are uncertain both because of uncertainties in cosmic abundances and because of uncertainties in measured gas-phase abundances.

\begin{figure}
\begin{center}
\includegraphics[scale=0.6]{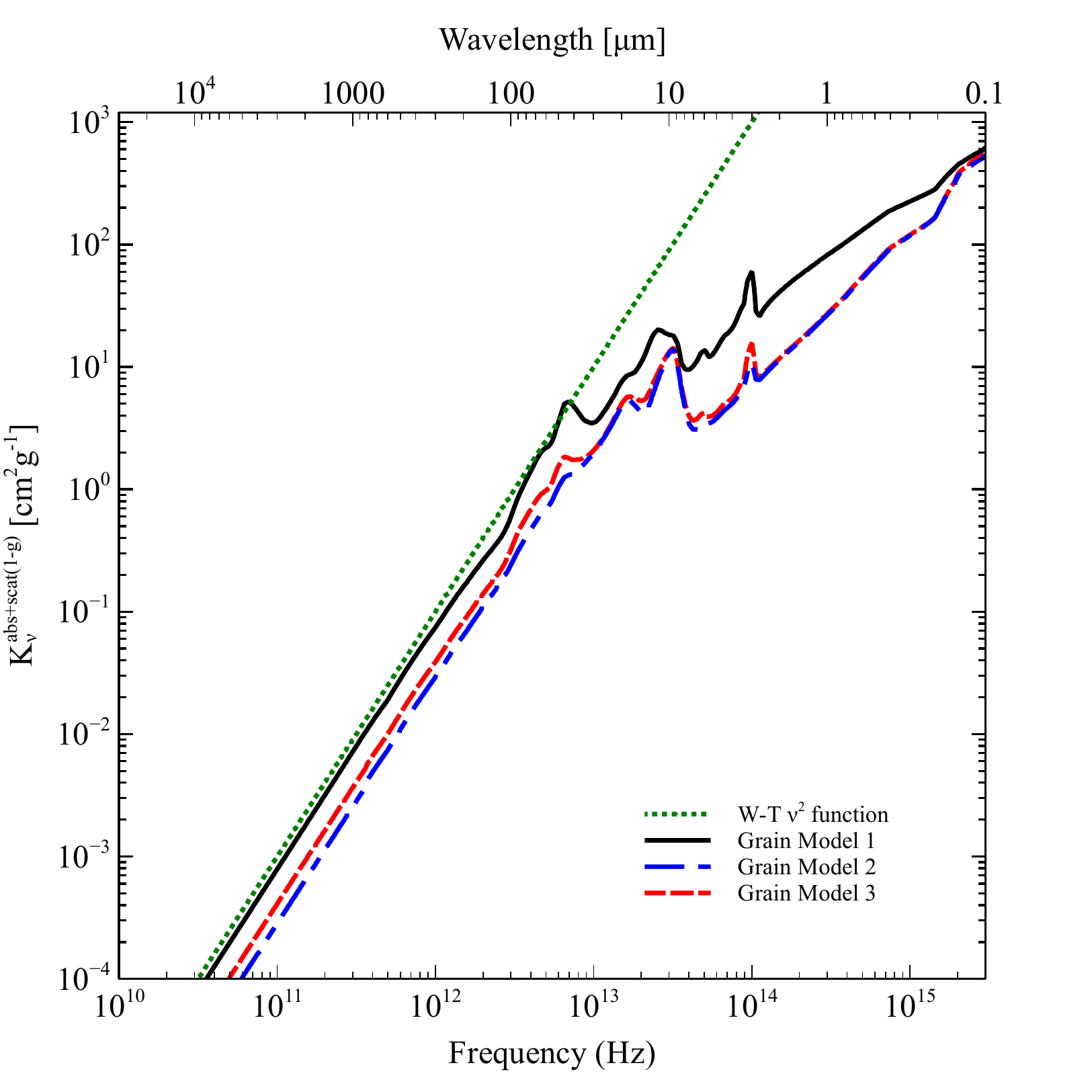}
\caption{
The gram opacity per unit mass of gas as a function of wavelength. The graph compares the opacities for our four grain models and \citetalias{WardTh10}’s assumption of \kn $\propto \nu^2$. All plots are the sum of absorption extinction plus scattering extinction, the latter corrected for forward scattering of radiation back into the line-of-sight to the observer. Grain model 1 is recreated using refractive index data provided by \citet{Preb93} with dirty ice mantle to silicate core ratio of b/a = 1.62. In grain models 2 and 3, we conserve the dust phase abundances for constituent elements and introduce a layer of vacuum between the the mantle and the core. Grain model 2 uses b/a = 1.62 and 70\% vacuum by volume, while grain model 3 uses b/a =2 , and 80\% vacuum by volume. The reasons for creating these models are further explained in the text. (A color version of this figure is available in the online journal.)}
\label{fig:opacity}
\end{center}
\end{figure}

Our grain models include many of the components discussed by \citetalias{Preb93}, especially core mantle particles (CMPs) that are expected to exist in cold molecular regions. As described by these authors, a CMP consists of a spherical silicate (generically, MgSiFeO$_4$) core surrounded by a “dirty ice” mantle. The dirty ice contains \water and \ammonia ices mixed with small amorphous carbon (aC) particles. \citetalias{Preb93} calculated refractive indices of dirty ice using EMT. They assumed a volume ratio of \water to \ammonia ices of 3:1, a ratio that best fits observations of the 3.1 $\micron$ ice feature toward the Becklin-Neugebauer (BN) object \citep[also, see][]{Hagen83}. \citetalias{Preb93} assumed that aC particles occupy 10\% of the dirty ice volume. We adopt these assumptions and incorporate the \citetalias{Preb93} dirty ice refractive indices into our grain models. Other parameters describing CMPs are (i) the ratio of mantle radius to core radius, b/a; (ii) the ratio of Si in grains to total H, Si/H; and (iii) the size distribution of the silicate core radii, n(a). Like \citetalias{Preb93}, we chose the MRN size distribution n(a) $\propto$ a$^{-3.5}$ \citep{Mathis77} for silicate cores in the size range listed in Table \ref{tab4}. The ratio b/a, and the assumed dirty ice composition described above, establish the ratios of C, N, O, Mg and Fe to Si in the CMPs. The Si/H ratio establishes the CMP abundances of all of these elements relative to H. The assumed core size distribution establishes the number of CMPs per unit H. The \citetalias{Preb93} grain models, like ours, also include free aC particles in addition to those within the dirty ice. As described above, a given assumed Si/H ratio implies a ratio C/H in the CMPs. If free aC particles are to be part of the overall grain model, then the total C/H ratio for grains must be greater than the C/H ratio in the CMPs alone.

We have constructed three different grain models with Cloudy, and we have calculated grain opacities for each. These models explore a range of grain properties, resulting in a range of calculated \kn. Each model assumes the MRN size distribution used by \citepalias{Preb93} for free aC particles and for CMP silicate cores (Table \ref{tab4}). Our calculations of \kn also use the same refractive index data for free aC particles, for silicates, and for dirty ice used by \citepalias{Preb93} in the wavelength range 0.1 to 800 $\micron$\footnote{We have extended the wavelength range of the aC, silicate and dirty ice refractive indices outside 0.1 to 800 $\micron$ on an \emph{ad hoc} basis. This extension, described in the Appendix, is done for computational compatibility with the Cloudy}. In addition, all of the grain models allocate nearly the same C/H ≈ 1\e{-4} to the free aC grains. Therefore, the contributions of free aC particles to the total opacities in all three models are essentially the same. The principal differences among the grain models arise from differences in properties of the CMPs, differences that imply different abundances of elements in grains and different values of \kn. We present the resulting values of \kn in Figure \ref{fig:opacity}. Opacities plotted in Figure \ref{fig:opacity} are the sum of absorption and scattering opacities, the latter corrected for forward scattering of photons from an extended background source into the line of sight of the observer\footnote{Scattering opacities for a point-like background source, such as a star, are higher than scattering opacities for an extended background source. This difference arises because a photon from a point source that is forward scattered by even a very small angle is not seen by the observer. In contrast, a photon from an extended source that is forward scattered by an angle no greater than the angular size of the background source is still seen by the observer. At FIR wavelengths, grain scattering is insignificant, so this distinction is irrelevant. However, scattering is important at optical wavelengths. As a result, point source opacities in the V-band are about 1.7 times greater than extended}. In addition, opacities in Figure \ref{fig:opacity} are the sums of opacities contributed by CMPs and by free aC grains. We refer readers to the appendix for a detailed description of each of these grain models.

Our grain model 1 is an attempt to replicate the \citetalias{Preb93} model that is the basis for the \citetalias{WardTh10} analysis of Herschel Polaris Flare data. This model provides a consistency check between Cloudy calculations of grain opacities and those of \citetalias{Preb93}. This model has b/a = 1.62, and abundance information from \citetalias{Preb93}. As expected, our model 1 yields \kn values in the FIR that are nearly identical to those calculated by \citetalias{Preb93} (see their figure 4, dotted line for b/a = 1.62). These values are also very close to the strict \kn $\propto \nu^2$ law used by \citetalias{WardTh10} in the FIR. This latter correspondence is shown in Figure \ref{fig:opacity}, where the solid line is \kn for model 1 and the straight dotted line is the $\nu^2$ law used by \citetalias{WardTh10}.

Based upon information in Table \ref{tab:appendix} and the relevant discussion in the appendix A, we conclude that model 1 uses the silicate core elements Mg, Si and Fe in near optimum abundances; however, it overuses the elements C, N, and O by factors 2, 6, and 2, respectively. Note that the overuse of C in model 1 could be eliminated by assuming all C in grains is within the CMPs, leaving none available for the free aC grains. However, the absence of free aC grains would affect the chemistry in the model, in particular, the predictions of CO line strengths.

Our second grain model explores the effect of a vacuum component. We replace some of the CMP mantle volume with a vacuum layer, thereby reducing the need for mantle elements C, N and O that are overused in model 1. For the reasons explained in the appendix A, we chose a vacuum layer between the silicate core and the dirty ice mantle. This choice introduces another free parameter, the fraction of the total CMP volume that is vacuum. For model 2 we chose 70\% vacuum, and we retained b/a = 1.62, as for model 1. Values of \kn for model 2 are plotted in Figure \ref{fig:opacity} (dashed-dotted blue line). Even if much of the mantle volume in model 1 has been replaced with vacuum in model 2, FIR opacities are nearly a factor of three less than opacities in model 1.

The third Cloudy grain model is an attempt to increase calculated values of \kn in the FIR while still using elements near their optimum abundances. To do so, we increased the CMP vacuum volume fraction to 80\%, and we increased b/a to 2.0 so that the CMPs are larger than in models 1 and 2. As shown in Table \ref{tab:appendix}, the required grain abundances are now very close to optimal for all six grain elements. (Compare Table\ref{tab:appendix}, columns (d) and (g).) Moreover, \kn plotted in Figure \ref{fig:opacity} (red dashed line) is higher in the FIR than for model 2, although still a factor of about two lower than opacities in model 1.

In summary, the grain models that do not overuse grain element abundances (models 2 and 3) both predict FIR grain opacities lower than those predicted by \citetalias{Preb93} and replicated in our grain model 1 (see table \ref{tab5}). Between models 2 and 3, the latter predicts higher FIR opacities although still a factor of two lower than \citetalias{Preb93}. We therefore considered the possibility of a grain model with b/a larger than model 3. Such a model would have larger size CMPs, a larger vacuum component, and, presumably, higher FIR opacities. However, the grain model would need to retain the same mantle volume as model 3 in order to require the same mantle element abundances that are optimal. This requirement implies a very thin mantle occupying only a very small fraction of the CMP volume, like an eggshell surrounding the yolk (the silicate core) with vacuum in between. We regard such a grain model as unphysical. We conclude that the FIR opacities of grain model 3 are the highest obtainable in vacuum grain models of the type discussed here that optimally use grain elements.

\begin{deluxetable}{ccccccc}
\tablecolumns{7}
\tablecaption{\label{tab5} Opacity Comparison}
\tabletypesize{\small}
\tablehead{
	\colhead{Grain Model} & \colhead{\kn(160)} & \colhead{\kn(350)} & \colhead{\kn(500)} &
		 \colhead{\kn(850)} & \colhead{\kn(450):\kn(2.2)} & \colhead{\kn(850):\kn(2.2)}\\
	\colhead{(a)} & \colhead{(b)} & \colhead{(c)} & \colhead{(d)} & \colhead{(e)} & \colhead{(f)} & \colhead{(g)}}
\startdata
WT10 & 0.349 & 0.073 & 0.036 & 0.012 & \nodata & \nodata \\
Model 1 & 0.234 & 0.056 & 0.028 & 0.010 & 9.6\e{-4} & 2.7\e{-4}\\
Model 2 & 0.096 & 0.022 & 0.011 & 0.003 & 1.3\e{-3} & 3.4\e{-4}\\
Model 3 & 0.124 & 0.023 & 0.015 & 0.005 & 1.7\e{-3} & 4.8\e{-4}\\
\enddata
\tablecomments{
The opacities \scmpg given here follow $\nu^2$ behavior in FIR range as shown in Figure \ref{fig:opacity}. These numbers can be compared with other predicted opacities in the literature \citep[e.g. Table 2 of][]{Shirley05}. These opacities are calculated per unit mass of gas. To convert these numbers to per unit mass of dust \citep[e.g. Table 2 of][]{Ormel11} multiply with the assumed gas to dust mass ratio.}
\end{deluxetable}

The dilemma over grain element abundances and observed optical grain opacities, as discussed in the appendix, leads us to propose our grain models 1 and 3 as useful extremes. Grain model 1, a replica of \citetalias{Preb93}, correctly predicts observed grain opacities in the optical V band, although it does so by overusing mantle grain elements in the CMPs by at least a factor of two. Model 3 is consistent with grain element abundances, although it under-predicts the observed optical V band opacities by at least a factor of two. In the FIR, opacities of model 1 are about twice those of model 3. In short, we have no grain model that meets all available observations of grains and grain element abundances. This situation may reflect limitations on our knowledge of grains for which the true structures and compositions are more complicated than existing grain models assume. Otherwise, the situation may reflect limitations on the numerical approximations used in calculating indices of refraction and grain opacities. If grain models 1 and 3 are useful extremes, we conclude that FIR grain opacities, hence, cloud column densities and cloud masses derived with them, are uncertain by at least a factor of two on these grounds alone. In particular, observers like \citetalias{WardTh10} who adopt the \citetalias{Preb93} FIR opacity law (replicated by our grain model 1) underestimate N(H) by about a factor of two if our grain model 3 is correct, instead. In effect, values of N(H) derived from the \citetalias{Preb93} opacity law may be considered lower limits.

\section{Effects of Dust Temperature Variations along the Line-of-sight}

\begin{figure}
\begin{center}
\includegraphics[scale=0.7]{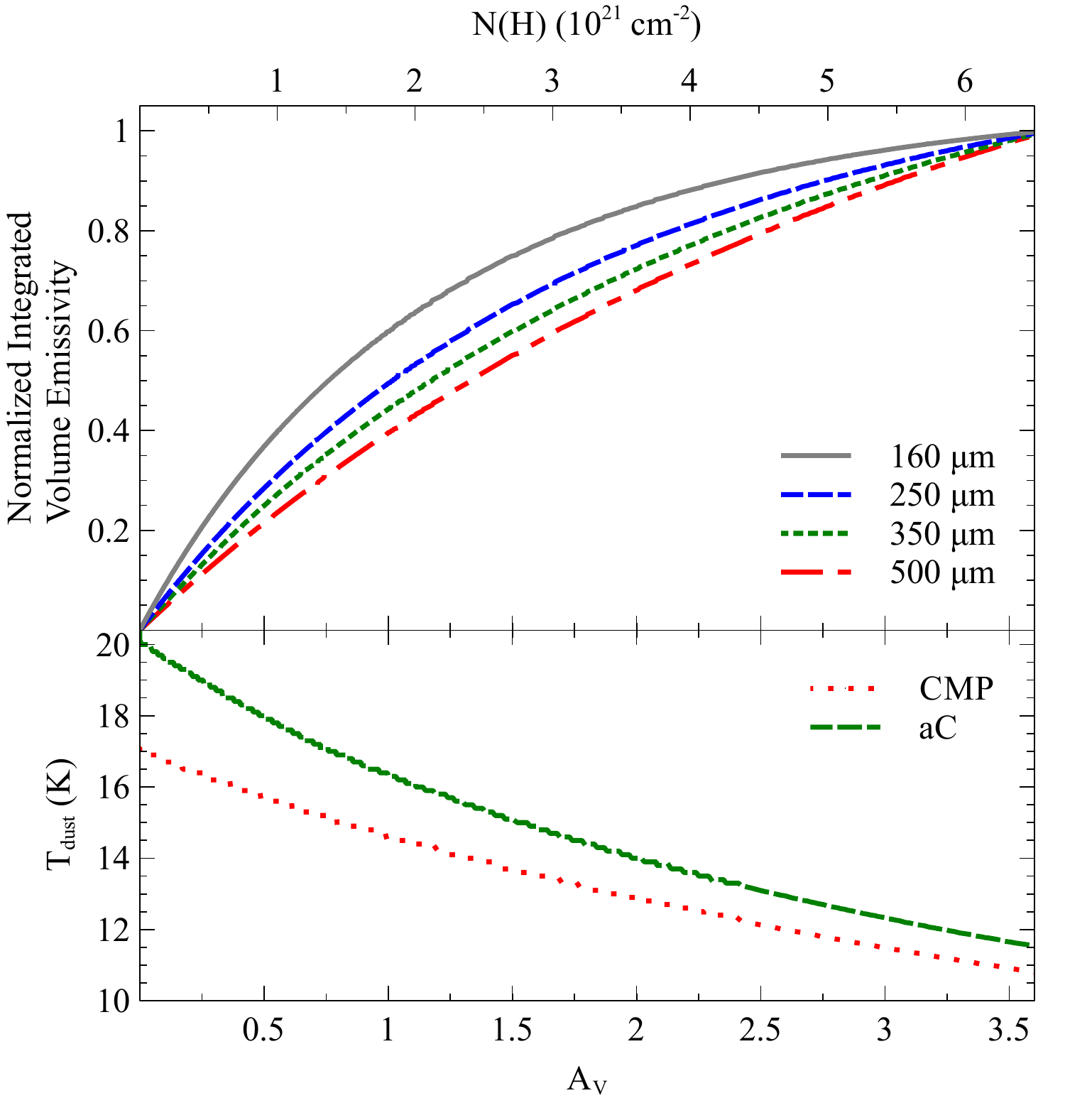}
\caption{
Normalized integrated volume emissivity in Herschel bands (top panel) and dust temperature (bottom panel) as a function of \Av for a model of core 1 with grain model 1. (A color version of this figure is available in the online journal.)}
\label{fig:td}
\end{center}
\end{figure}

Uncertainties in N(H) derived from FIR observations can arise not only because of uncertainties in FIR grain opacities (\S5); uncertainties can also arise from variations in \Td along a given line-of-sight. Observers commonly derive N(H) by fitting a modified blackbody function to observed FIR brightnesses in several bands \citepalias[e.g.][; also see \S2]{WardTh10}. The fitting process returns estimates of N(H) and \Td, the latter assumed to be constant along the line-of-sight. However, \Td declines with depth into an externally heated core, and FIR emissivities are a strong function of \Td. For example, at \Td $\approx$ 15 K, the 250 $\micron$ emissivity scales approximately as (\Td)$^4$. As a result, most of the emergent FIR radiation from a core may arise from the warmer outer layers of the core. Therefore, the observed FIR brightnesses are not linearly proportional to N(H) even though the core is optically thin in the FIR. To illustrate this effect, we present model calculations in Figure \ref{fig:td} for MC123 core 1 (with grain model 1). The bottom panel of this figure shows the decline of \Td into the cloud, with \Td plotted separately for CMPs and free aC particles. The top panel shows the normalized integrated volume emissivities for the various Herschel bands. These results are plotted as a function of Av and N(H). Judging from Figure \ref{fig:td}, the outer layer of the model cloud (\Av $<$ 2 mag, N(H) $<$ 3.5\e{21} \pscm) contributes 60-80\% of FIR emission, depending upon wavelength. This result implies that FIR observations of clouds with \Av $\gg$ 2 mag are not particularly sensitive to the total N(H) because the cold, inner regions of the clouds contribute very little emission. Therefore, values of N(H) for such clouds derived from FIR observations may be underestimates.

We use Cloudy models to explore the sensitivity of FIR dust emission to increasing N(H) in molecular clouds. In particular, we study the accuracy of the modified blackbody fitting technique used by observers. Our study is possible because Cloudy predicts FIR brightnesses in the Herschel bands for a model cloud of specified N(H)$_{\rm Cloudy}$. The predicted brightnesses can then be fitted to the modified blackbody function used by observers to determine N(H)$_{\rm fit}$. Finally, N(H)$_{\rm fit}$ is compared with N(H)$_{\rm Cloudy}$, the latter taken as the “true” N(H).

\begin{figure}
\begin{center}
\includegraphics[scale=0.55]{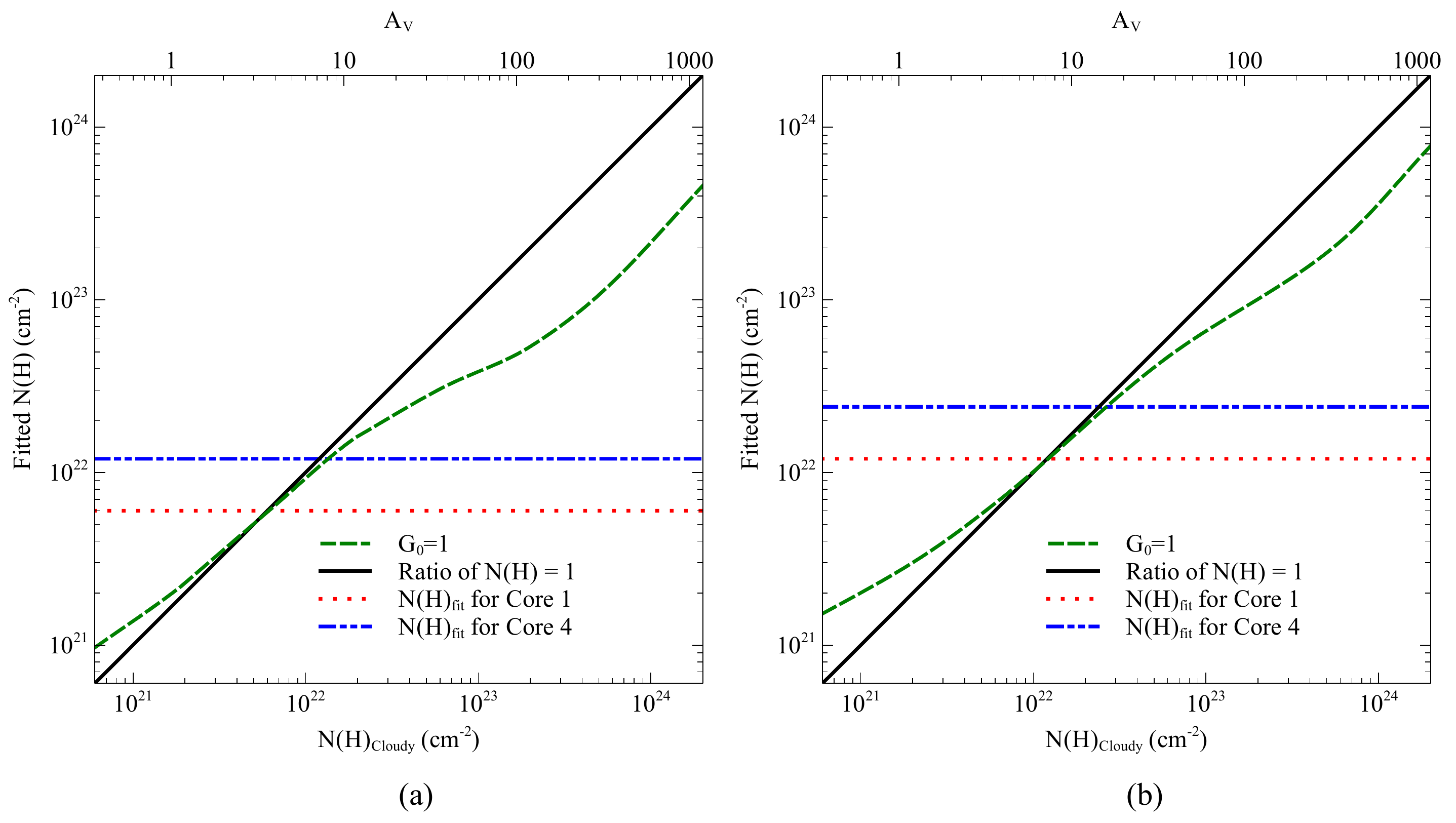}
\caption{
The comparison between N(H)$_{\rm Cloudy}$ and N(H)$_{\rm fit}$ for Cloudy models with G$_0$ = 1. The graph on left is for grain model 1, and one on the right is for our grain model 3. The solid line represents N(H)$_{\rm fit}$ = N(H)$_{\rm Cloudy}$. The horizontal dotted and dash-dotted lines correspond to core 1 and core 4 column densities fitted as described in text. Note that the two grain models have different N(H)/\Av ratios. Also, N(H)fit for cores is different for the two grain models by a factor of about 2, reflecting the difference in opacities at 1000 GHz between the two models. (A color version of this figure is available in the online journal.)}
\label{fig:6}
\end{center}
\end{figure}

We created a series of Cloudy models with N(H)$_{\rm Cloudy}$ varying from 6\e{20} \scm to 2\e{24} \scm in steps of $\sim$0.5 dex. For these models, we used the density law for MC123 core 1, increasing N(H)$_{\rm Cloudy}$ by increasing d$_{\rm stop}$ alone. All models used for this first study have G$_0$ = 1. Also, values of N(H)$_{\rm fit}$ were in all cases determined by fits to Cloudy-predicted brightnesses in the Herschel 160, 250, 350 and 500 $\micron$ bands. Figure \ref{fig:6} shows comparisons between N(H)$_{\rm fit}$ and N(H)$_{\rm Cloudy}$ using grain model 1 (Figure \ref{fig:6}a) and 3 (Figure \ref{fig:6}b). The diagonal straight line in each plot represents N(H)$_{\rm fit}$ = N(H)$_{\rm Cloudy}$. The horizontal lines denote values of N(H)$_{\rm fit}$. In Figure \ref{fig:6}a, the horizontal lines represent N(H)$_{\rm fit}$ taken from Table \ref{tab2} for M123 core 1 (lower line) and core 4 (upper line). In Figure \ref{fig:6}b, the horizontal lines are analogous, except they represent N(H)$_{\rm fit}$ for the same two cores if grain model 3 rather than grain model 1 is used. The horizontal axis of each plot is labeled in N(H)$_{\rm Cloudy}$ and in the equivalent \Av as determined by Cloudy. (The ratio \Av/N(H) is a factor of 1.7 higher for grain model 1 than grain model 3.) With both grain models (Figure \ref{fig:6}, dashed green lines), N$_{\rm fit}$ $\sim$ N$_{\rm fit}$ for \Av $\sim$ 1--10 mag (i.e. N(H)$_{\rm Cloudy}$ $\sim$ 2\e{21} -- 2\e{22} \pscm). This result suggests that the modified blackbody fitting technique with constant \Td yields accurate (within $\sim$ 20\%) column densities in this range of \Av or N(H). This range includes all MC123 cores. For both grain models, however, N$_{\rm fit}$ $<$ N$_{\rm Cloudy}$ for \Av $>$ 10 mag (i.e. N(H)$_{\rm Cloudy}$ $>$ 2\e{22} \pscm). If Cloudy models resemble real molecular cores, then use of the modified blackbody fit for Av $\gg$ 10 underestimates N(H) by factors of up to 5 and 3 for grain models 1 and 3, respectively. Note that these potential underestimates of N(H) from the modified blackbody fitting process are in addition to underestimates of N(H) that can arise from uncertainties in FIR grain opacities (see \S4).

Conclusions drawn above about possible underestimates of N(H) are based upon the assumption that G$_0$ = 1. We now consider two related questions: is the assumption that G$_0$ = 1 likely to be correct, and are the results just described sensitive to the assumed value of G$_0$? To investigate the first of these two questions, we constructed a series of models that vary in G$_0$. Each model is based upon the density law of MC123 core 1. One set of models used grain model 1 while the other set used grain model 3. For each Cloudy model and for each of the four Herschel bands, we calculated the ratio of the Cloudy-predicted FIR brightness to the observed brightness for core 1 as listed in Table \ref{tab2}. We also calculated a goodness of fit parameter for each Cloudy model, $\chi^2$, where
\begin{equation}
\chi^2 = \sum_{}^{}{\left(\frac{Predicted Flux}{Observed Flux} - 1\right)}^2
\end{equation}

\begin{deluxetable}{cccccc}
\tablecolumns{6}
\tablecaption{\label{tab6} Comparison of Predictions for Dust Emission in core 1 with Variation in G$_0$ for Grain Model 1}
\tabletypesize{\small}
\tablewidth{0pt}
\tablehead{
	\colhead{Incident Field} & \multicolumn{4}{c}{Herschel Bands} & \colhead{$\chi^2$}\\
	\colhead{G$_0$} & \colhead{160 $\micron$} & \colhead{250 $\micron$} & \colhead{350 $\micron$} & 
		\colhead{500 $\micron$} & \colhead{}\\
	\colhead{(a)} & \colhead{(b)} & \colhead{(c)} & \colhead{(d)} & \colhead{(e)} & \colhead{(f)}}
\startdata
0.4 & 0.5 & 0.6 & 0.8 & 1.0 & 0.49\\
0.6 & 0.8 & 0.8 & 1.1 & 1.2 & 0.14\\
0.8 & 1.0 & 1.0 & 1.3 & 1.4 & 0.24\\
1.0 & 1.3 & 1.2 & 1.5 & 1.6 & 0.68\\
1.2 & 1.6 & 1.4 & 1.7 & 1.7 & 1.42\\
1.4 & 1.9 & 1.6 & 1.8 & 1.8 & 2.42\\
1.6 & 2.1 & 1.7 & 2.0 & 2.0 & 3.64\\
\enddata
\tablecomments{
Values in columns (b) through (e) are the ratios of Cloudy predicted brightness to the observed brightness 
for each Herschel band.}
\end{deluxetable}

Table \ref{tab6} presents the ratios of Cloudy-predicted to observed FIR brightnesses for core 1 and grain model 1, as well as $\chi^2$ values. Table \ref{tab7} presents the same information for core 1 using grain model 3. The $\chi^2$ parameter is minimized with grain model 1 (Table \ref{tab6}) for G$_0 \approx$ 0.7; the $\chi^2$ parameter is minimized with grain model 3 (Table \ref{tab7}) for G$_0 \approx$ 1.1. That is, Cloudy models of Polaris Flare core 1 best fit the observations in four Herschel bands when G$_0 \approx$ 1. We take this result as an indication that the assumption of G$_0 \approx$ 1 is reasonable for models of the Polaris Flare and, by extension, for models of similar cold, starless cores in regions without an enhanced ISRF. Note that the slightly different best fit values for G$_0 \approx$ using grain models 1 and 3 is expected. Grain model 3 yields lower FIR opacities than grain model 1, as previously noted. Therefore, to predict the same FIR brightness in a given band with grain model 3, slightly higher vales of \Td are necessary. Higher values of \Td, of course, are produced by higher values of G$_0$.

\begin{deluxetable}{cccccc}
\tablecolumns{6}
\tablecaption{\label{tab7} Comparison of Predictions for Dust Emission in core 1 with Variation in G$_0$ for Grain Model 3}
\tabletypesize{\small}
\tablewidth{0pt}
\tablehead{
	\colhead{Incident Field} & \multicolumn{4}{c}{Herschel Bands} & \colhead{$\chi^2$}\\
	\colhead{G$_0$} & \colhead{160 $\micron$} & \colhead{250 $\micron$} & \colhead{350 $\micron$} & 
		\colhead{500 $\micron$} & \colhead{}\\
	\colhead{(a)} & \colhead{(b)} & \colhead{(c)} & \colhead{(d)} & \colhead{(e)} & \colhead{(f)}}
\startdata
0.4 & 0.4 & 0.4 & 0.5 & 0.6 & 1.13\\
0.6 & 0.6 & 0.5 & 0.7 & 0.7 & 0.52\\
0.8 & 0.8 & 0.7 & 0.9 & 0.8 & 0.18\\
1.0 & 1.1 & 0.8 & 1.0 & 0.9 & 0.04\\
1.2 & 1.3 & 0.9 & 1.1 & 1.0 & 0.08\\
1.4 & 1.5 & 1.1 & 1.2 & 1.1 & 0.26\\
1.6 & 1.7 & 1.2 & 1.3 & 1.2 & 0.58\\
\enddata
\tablecomments{
Values in columns (b) through (e) are the ratios of Cloudy predicted brightness to the observed brightness 
for each Herschel band.}
\end{deluxetable}

\begin{figure}
\begin{center}
\includegraphics[scale=0.55]{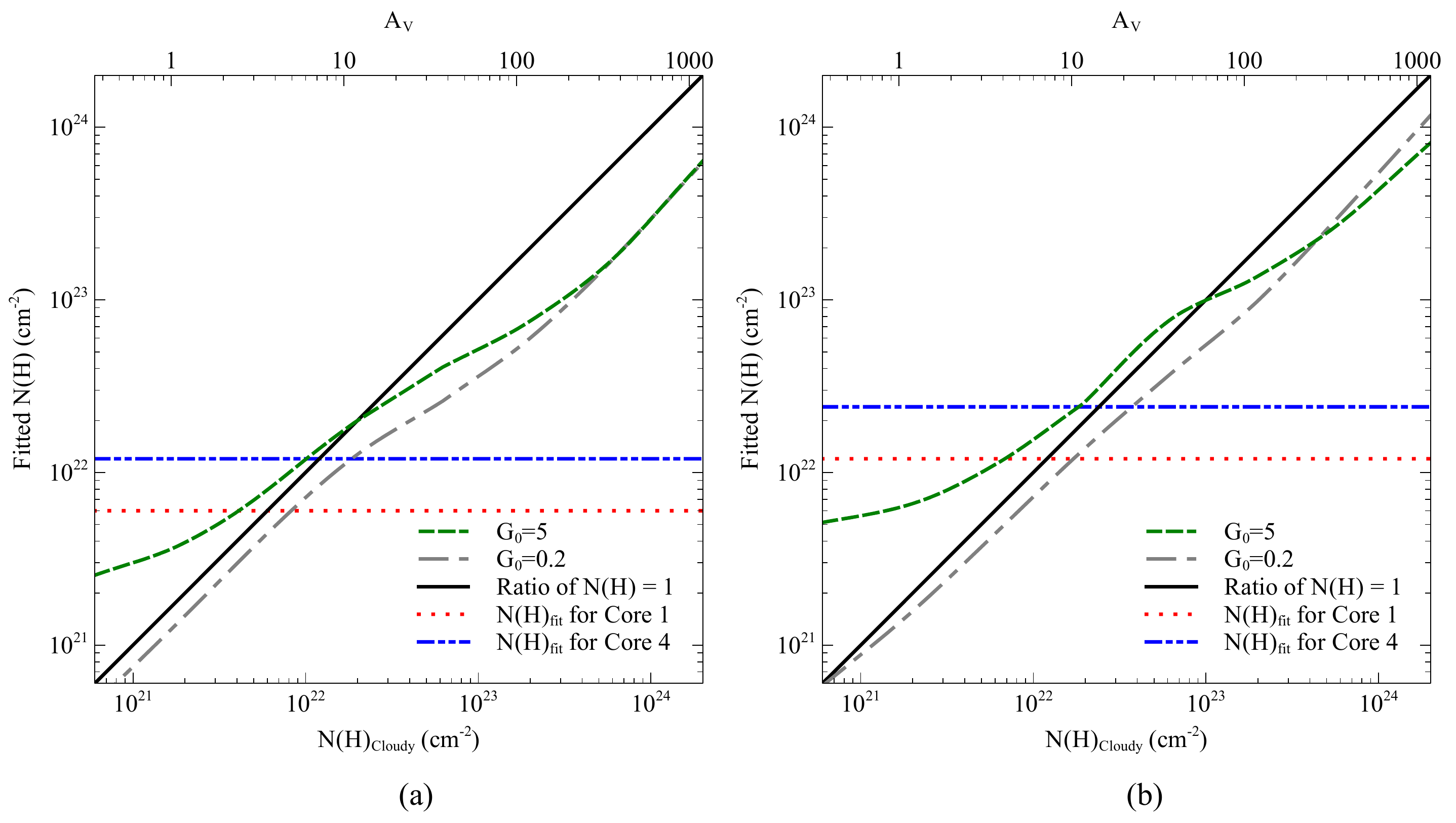}
\caption{
The comparison between N(H)$_{\rm Cloudy}$ and N(H)$_{\rm fit}$ for Cloudy models with G$_0$ = 0.2 and 5, with grain model 1 (a), and grain model 3 (b). Also see caption for Figure \ref{fig:6}. (A color version of this figure is available in the online journal.)}
\label{fig:7}
\end{center}
\end{figure}

We now consider the question of whether potential underestimates of N(H) with the standard modified blackbody fitting technique (Figure \ref{fig:6} and related discussion) are sensitive to G$_0$. Even if G$_0 \approx$ 1 in the Polaris Flare region, other cold, starless cores might reside in environments of somewhat stronger or weaker ISRF. To explore this issue, we computed sets of Cloudy models similar to those used to construct Figure \ref{fig:6} but with G$_0$ = 0.2 and G$_0$ = 5. As for Figure \ref{fig:6}, we used both grain models 1 and 3. In Figure \ref{fig:7} we present comparisons of N(H)$_{\rm fit}$ and N(H)$_{\rm Cloudy}$ for Cloudy models with G$_0$ = 0.2 and G$_0$ = 5. Figure \ref{fig:7}a presents results with grain model 1; Figure \ref{fig:7}b presents results with grain model 3. Horizontal lines in Figure \ref{fig:7} are the same as those in Figure \ref{fig:7}. The results in Figure \ref{fig:7} show some sensitivity to G$_0$. For example, at low \Av (1--10 mag), N$_{\rm fit} >$ N$_{\rm Cloudy}$ for G$_0$ = 5, while N$_{\rm fit} <$ N$_{\rm Cloudy}$ for G$_0$ = 0.2. This statement holds for Cloudy models using both grain models. Evidently, the modified blackbody fitting technique can somewhat overestimate N(H) for G$_0 \gg$ 1 and underestimate N(H) for G$_0 \ll$ 1. However, these effects are relatively modest (less than a factor of 2) even over the relatively large (25:1) range in G$_0$ considered here. For \Av $>$ 10, especially \Av $\gg$ 10, we still find N$_{\rm fit} <$ N$_{\rm Cloudy}$ for G$_0$ over the full range in G$_0$ and for both grain models. We conclude that our previous statements regarding N$_{\rm fit}$ and N$_{\rm Cloudy}$ (based upon G$_0$ = 1 models) are approximately valid, especially if G$_0$ does not deviate significantly from the average interstellar value of $\approx$ 1. That is, the standard observers’ modified blackbody fitting technique is accurate in estimating N(H) to better than a factor of 2 for clouds with \Av $<$ 10. For \Av $\gg$ 10, the fitting technique typically underestimates N(H) by a factor of 2--5, depending upon the value of G$_0$. Given the range of possible grain models and values of G0, a typical underestimate of N(H) for clouds with Av $gg$ 10 is of order a factor of 3. This latter conclusion is especially relevant to high column density molecular clouds such as IRDCs.

\section{Implications for Virial Stability of Starless Molecular Cores}

If starless cores are to form stars eventually, their masses must be comparable to or greater than their virial masses. We investigated this question with specific reference to the MC123 cores for which FIR Herschel and other data exist. For each core, we calculated the mass M$_{\rm cor}$ using N(H) from a modified blackbody fit with grain model 1, see values in Table \ref{tab2}. (Use of grain model 3 would result in higher N(H), hence, higher masses by about a factor of 2, see \S4.)  For each core we also calculated the virial mass M$_{\rm vir}$, with radii taken from Table \ref{tab2} and $\Delta v$, the FWHM velocity width of the \thco line, taken from the data of \citet{HBF}. We calculated virial masses using equation 3 of \citet{MacLaren88} with their constant \emph{k$_2$} = 126, the value appropriate to a core with an r$^{‒2}$ density law. (For a constant density core, \emph{k$_2$} = 210, so virial masses are 1.67 times higher.) In Table \ref{tab8}, we present values of $\Delta v$ from the \thco lines towards each core as well as values of M$_{vir}$ and the ratios M$_{\rm vir}$/M$_{\rm cor}$. Clearly,  M$_{\rm vir}$/M$_{\rm cor} \gg$ 1 for all cores, especially cores 1 through 3. This conclusion holds even if we adopt masses based upon grain model 3 (i.e. M$_{\rm cor}$ values about a factor of 2 higher), especially since we have used the lowest likely value of \emph{k$_2$} in calculating M$_{\rm cor}$. In short, our estimates of M$_{\rm vir}$/M$_{\rm cor}$ strongly suggest that the M123 cores are not self-gravitating. If so, then they must either be pressure confined or else unstable to expansion. However, pressure confinement of the dense cores seems unlikely unless much warmer gas surrounds the MC123 region. The MC123 cores may well be transient features of turbulence in the Polaris Flare cloud, not the sites of future star formation.

\begin{deluxetable}{cccccc}
\tablecolumns{6}
\tablecaption{\label{tab8} Virial masses for cores from \thco line-widths}
\tablewidth{0pt}
\tabletypesize{\small}
\tablehead{
	\colhead{} & \colhead{Core 1} & \colhead{Core 2} & \colhead{Core 3} & \colhead{core 4} & \colhead{Core 5}\\
	\colhead{(a)} & \colhead{(b)} & \colhead{(c)} & \colhead{(d)} & \colhead{(e)} & \colhead{(f)}}
\startdata
$\Delta v_{\,^{13}\rm CO}\,(\!\kmps\!\!)$ & 1.5 & 1.4 & 1.8 & 1.0 & 0.7\\
M$_{\rm vir}$ (M$_\sun$) & 7 & 9 & 18 & 4 & 2\\
M$_{\rm vir}$/M$_{\rm cor}$ & 53 & 27 & 31 & 8 & 5\\
\enddata
\end{deluxetable}

\citetalias{WardTh10} also considered the virial stability of the MC123 cores, also finding that  M$_{\rm vir}$/M$_{\rm cor} >$ 1, with values in the range 2-4. However, they hesitated to declare the cores non self-gravitating, given the inevitable uncertainties in mass calculations. Our much larger ratios  M$_{\rm vir}$/M$_{\rm cor}$ (Table \ref{tab8}) compared to those of \citetalias{WardTh10} are the result of our use of much larger values of $\Delta v$. WT10 assumed $\Delta v \approx$ 0.2 -- 0.4 \kmps, based upon aperture synthesis observations of the cores \citepalias[See references in][]{WardTh10}. In contrast, the single dish \thco data of \citet{HBF} yield $\Delta v \approx$ 1.0 -- 1.5 \kmps (Table \ref{tab8}), hence, much larger values of M$_{\rm vir} \propto \Delta v^2$. We find the single dish \thco data much more suitable for calculations of M$_{\rm vir}$ since the spatial resolution ($\approx$ 23 $\arcsec$) is roughly comparable to the sizes of the MC123 cores, and single dish observations do not suffer from missing flux. In contrast, aperture synthesis observations, with their much higher spatial resolutions, emphasize emission from small regions within the cores where the velocity structure is likely to be more quiescent than in the cores as a whole.

Of course, sampling issues complicate any such virial analysis. Ideally, one would establish $\Delta v$ from an optically thin spectral line whose emissivity is proportional to the FIR emissivity used to estimate N(H), hence, mass.  However, the emissivities of spectral lines and that of FIR continuum emission scale differently with density.  So the ideal cannot be met. \citet{Heith02} arrive at a different conclusion based on virial estimates using HC$_{\rm 3}$N lines. \citeauthor{Heith02} emphasize that their dust continuum emission map has morphology very similar to that of the previously published C$^{\rm 18}$O emission map. They also find that the HC$_{\rm 3}$N morphology is quite different from that of the dust continuum and CO. \thco should closely track C$^{\rm 18}$O, even if the former is slightly optically thick. On the other hand, HC$_{\rm 3}$N may be a tracer of regions in the cores where the chemistry is favorable to HC$_{\rm 3}$N formation. So their conclusions drawn from HC$_{\rm 3}$N observations are not comparable to those we draw from CO and dust emission. Perhaps, they are appropriate to differently sampled regions of the core. Also, there is the phenomenon of freeze-out of molecules which further complicates the sampling issue.  Nonetheless, we believe that values of $\Delta v$ inferred from single-dish \thco data are the most appropriate to use in this virial analysis. Our conclusion follows from a comparison between integrated \thco line intensities \citet{HBF} and FIR brightnesses across MC123.  We find that the morphologies of the two tracers are very similar.  More quantitatively, we calculated the contrast ratios of several MC123 cores in \thco and in 350 $\micron$ emission.  (The contrast ratio is the ratio between the peak brightness and the brightness at a nearby position away from the core.)  We chose 350 $\micron$ emission because the spatial resolution of this map (Figure \ref{fig:maps}, Table \ref{tab1}) is nearly identical to that of the \thco line maps of \citet{HBF}.  We find that the contrast ratios in \thco are one half to two thirds those in 350 $\micron$ emission.  This result suggests that \thco is slightly optically thick. Therefore, the \thco line widths may be slightly widened by optical depth effects or, equivalently, the \thco lines may weight the outer regions of a given core somewhat more than the inner regions.  However, this effect is likely to be minimal, especially in comparison with the strong weighting effect of aperture synthesis spectral line observations, which strongly favors the densest, innermost regions of the cores.  In short, \thco data appear to provide the most reliable available indicators of $\Delta v$ for the MC123 cores observed in the FIR.  Even if modest \thco optical depth effects increase the measured $\Delta v$ slightly, this increase is unlikely to alter the conclusions above about the virial stability of the core since  M$_{\rm vir}$/M$_{\rm cor} \gg$ 1.

\section{Summary and Conclusions}

We consider the relationship between observed FIR dust emission and N(H) in starless cores, using cores within the Polaris Flare as examples. To facilitate this study, we combine two sets of Herschel Space Telescope imaging data for cloud MCLD 123.5+24.9 of the Polaris Flare (referred to as MC123). We use the latest calibration pipeline, and we present images of MC123 in the Herschel bands centered at 160 $\micron$, 250 $\micron$, 350 $\micron$ and 500 $\micron$. These images contain five starless cores.

We use the Cloudy, a spectral synthesis code, to investigate the relationship between N(H) and FIR emission and its uncertainties. Cloudy can calculate dust opacities from models of interstellar grains. Given such a grain model, Cloudy can calculate molecular core models in which physical conditions, including dust temperatures and grain emissivities, are followed as a function of depth into the core until a specified N(H)$_{\rm Cloudy}$ is reached. Cloudy then predicts the emergent radiation over a wide range of wavelengths, including FIR wavelengths to which the Herschel Space Telescope is sensitive. The starless cores are assumed to be externally heated by the interstellar radiation field (ISRF) that, itself, can be varied in the models.

From this study, we draw the following principal conclusions:

\begin{enumerate}
	\item{FIR grain opacities, which cannot be directly measured, are uncertain by at least a factor of two, leading to uncertainties in derived N(H) by the same factor. We focus upon two Cloudy grain models, each of which consists of core mantle particles (CMPs) and free amorphous carbon (aC) particles, with specified elemental abundances and size distributions. Grain model 1 is essentially the same as that described by \citetalias{Preb93} and used by \citetalias{WardTh10} and other observers to derive values of N(H) from observations of FIR dust emission. However, this model requires higher abundances of C, N and O than observations suggest for the ISM. Grain model 3 (model 2 is used only as an illustrative example) incorporates a vacuum layer into the CMPs. The vacuum layer reduces the required ISM elemental abundances to those implied by observations. However, the calculated FIR opacities of grain model 3 are only about on half those of grain model 1. Therefore, use of grain model 3 to derive N(H) will result in N(H) (hence, core masses) that are twice as large as with grain model 1.}
	
	\item{Observers commonly fit a modified blackbody function to FIR observations over a range of wavelengths to derive N(H) and the dust temperature \Td along a given line-of-sight. This technique assumes a constant \Td. However, \Td (hence, dust emissivity) must decline with depth into an externally heated core. As a result, much of the observed FIR dust emission from high-N(H) cores should come from the outer layers of the core. Hence, the modified blackbody fitting technique may not be sensitive to dust in the colder interior of the core, and estimates of N(H) may be too low. We investigate this possible source of error with a series of Cloudy core models of increasing N(H)Cloudy. For each model, we fit the Cloudy-predicted FIR emission to a modified blackbody function, mimicking the technique of observers. We then compare the fitted column density N(H)$_{\rm fit}$ with the “true” column density N(H)$_{\rm Cloudy}$. We find that for N(H)Cloudy in the range $\approx$ 2\e{21} -- 2\e{22} \pscm (\Av $\approx$ 1--10 mag), N(H)$_{\rm fit} \approx$ N(H)$_{\rm Cloudy}$, that is, the observers’ fitting technique returns nearly the true value of N(H) for the molecular core.  This conclusion applies to the Polaris Flare cores. However, for N(H)$_{\rm Cloudy} \gg$ 2\e{22} \pscm (\Av $\gg$ 10 mag), we find that N(H)$_{\rm fit}$ underestimates N(H)$_{\rm Cloudy}$ by a factor of typically 3. This potential underestimation of N(H) by a factor of order 3 may be important to the study of high-N(H) molecular cores such as those in IRDCs.}
	
	\item{We consider the virial stability of the five starless cores in the Polaris Flare. We find that their masses, derived from FIR dust emission, are much less (5 -- 50 times less) than their virial masses, based upon \thco line widths from \citet{HBF}. Therefore, these cores are unstable to expansion and likely to be transient structures, perhaps the result of turbulent processes in the Polaris Flare. In short, there may be a very good reason why these cores are starless.}

\end{enumerate}

\acknowledgements
\emph{Acknowdgements --} We would like to thank the anonymous referee for the thoughtful review, which helped us improve the quality of this article. We are grateful to Babar Ali, Bernhard Schulz, and Dave Shupe at NASA Herschel Science Center help desk for their help and support, and for providing a virtual machine during the processing of the data in HSA. We would also like to thank Peter van Hoof for his help regarding the grain models. GJF acknowledges support by NSF (1108928, 1109061, and 1412155), NASA (10-ATP10-0053, 10-ADAP10-0073, NNX12AH73G, and ATP13-0153), and STScI (HST-AR- 13245, GO-12560, HST-GO-12309, GO-13310.002-A, and HST-AR-13914), and to the Leverhulme Trust for support via the award of a Visiting Professorship at Queen’s University Belfast (VP1-2012-025).

\appendix

\section{Grain Models}

In this section, we discuss in details the three grain models that we used in our analysis. We also discuss some of the computational details of the grain models in Cloudy. 

(1) Grain model 1 is based on \citetalias{Preb93} model that has b/a = 1.62 and Si/H = 3.1\e{5}. The model also assumes C/H = 2.2\e{-4} for CMPs and free aC grains combined, a value reflecting abundance information cited by \citetalias{Preb93}. This ratio allows 54\% of the aC in grains to reside outside the mantles as free aC grains, with the remaining aC in the CMP mantles. As expected, FIR opacities calculated for this model are nearly identical to those calculated by \citetalias{Preb93} and very close to the strict \kn $\propto \nu^2$ law used by \citetalias{WardTh10} in the FIR. Despite these similarities between our model 1 and the \citetalias{Preb93} model, we note one difference. Contributions to the opacity from free aC particles in our model 1 (not shown separately in Figure \ref{fig:opacity}) are about six times higher than those in the \citetalias{Preb93} model over the approximate wavelength range 5 -- 200 $\micron$ \citepalias[see figure 3 of][]{Preb93}. We have used the same aC refractive index data as \citetalias{Preb93} and standard Mie theory. Therefore, it is unclear why these differences exist in calculated aC particle opacities. However, the differences have no significant effect upon the calculated FIR opacities in model 1 because the CMPs, not the aC particles, strongly dominate the opacity at these wavelengths.

In table \ref{tab:appendix}, we present abundance data for the six grain elements considered in the models. Column (b) in this table lists gas phase abundances for the highly depleted ISM based on the compilation of \citet{Jenkins09}. Column (c) lists solar abundances from \citet{Asplund09} and column (d) has the differences (c) -- (b), that is, the abundances presumed to be in grains. Note that C/H in column (d) is half the value (2.2\e{-4}) taken by \citetalias{Preb93}, reflecting more recent estimates of cosmic abundances and depletions for C. The element abundances required by grain model 1 listed in column (e) are in excess of the abundances available for grains in column (d) for C, N and O by factors of 2, 6, and 2, respectively. In this sense, these three elements are overused by the grain model. Note that the overuse of C in model 1 could be eliminated by assuming all C in grains is within the CMPs, leaving none available for the free aC grains. However, the absence of free aC grains would affect the chemistry in the model, in particular, the predictions of CO line strengths. Grain model 1 uses elements Mg, Si and Fe in just about the optimal abundances.

The overuse of N in model 1 by a factor of six raises several issues. As indicated above, N is very weakly depleted in the ISM, if at all. \citet{Jenkins14} lists the gas-phase abundance of N as between 60\% and 100\% of the solar abundance, leaving no more than 20 $\pm$ 20\% of the solar N abundance available for grains. Yet model 1 requires 95\% of the solar N abundance in grains \citep[and 125\% of the B star N abundance in grains as listed by][]{Asplund09}. If \ammonia ice is indeed present in dirty ice mantles in the assumed proportion, then model 1 is ruled out on N abundance grounds. However, the evidence for \ammonia ices in CMPs is far from conclusive. As mentioned earlier, the inclusion of \ammonia ice in the \citetalias{Preb93} model (hence, in our model 1) is based upon an argument by \citet{Hagen83} about the origin of the 3.1 $\micron$ ice feature in the spectrum of the BN object. This argument may not apply to the much cooler environments of the Polaris Flare cores and cores like them. Therefore, it is entirely possible that less N (or even none at all) exists in the CMPs in these regions, implying mantles of pure or nearly pure \water ice. In this case, the overuse of N in grain models becomes moot. We have not further considered this possibility in our grain models.

\begin{deluxetable}{ccccccc}
\tablecolumns{7}
\tablenum{A.1}
\tablecaption{\label{tab:appendix} The Composition of the Grains}
\tabletypesize{\small}
\tablehead{
	\colhead{Element} & \colhead{Highly} & \colhead{Solar} & \colhead{Available} &
		 \colhead{Grain} & \colhead{Grain} & \colhead{Grain}\\
	\colhead{} & \colhead{-depleted ISM$^{1}$} & \colhead{Abundance$^{2}$} & \colhead{Grain Abundance} &
		 \colhead{Model 1$^{3}$} & \colhead{Model 2} & \colhead{Model 3}\\
	\colhead{(a)} & \colhead{(b)} & \colhead{(c)} & \colhead{(d)} & \colhead{(e)} & \colhead{(f)} & \colhead{(g)}}
\startdata
C & 1.9\e{-4} & 3.0\e{-4} & 1.1\e{-4} & 2.2\e{-4} & 1.1\e{-4} & 1.1\e{-4}\\
N & 6.2\e{-5} & 7.4\e{-5} & 1.2\e{-4} & 7.0\e{-5} & 5.8\e{-6} & 1.2\e{-5}\\
O & 3.3\e{-4} & 5.3\e{-4} & 2.0\e{-4} & 3.6\e{-4} & 1.5\e{-4} & 1.7\e{-4}\\
Mg & 2.3\e{-6} & 4.4\e{-5} & 4.1\e{-5} & 3.1\e{-5} & 3.3\e{-5} & 3.3\e{-5}\\
Si & 1.8\e{-6} & 3.5\e{-5} & 3.3\e{-5} & 3.1\e{-5} & 3.3\e{-5} & 3.3\e{-5}\\
Fe & 2.0\e{-7} & 3.5\e{-5} & 3.3\e{-5} & 3.1\e{-5} & 3.3\e{-5} & 3.3\e{-5}\\
\enddata
\tablecomments{
Column (d) lists the maximum available grain abundance for each element, that is, column (c) minus column (b). Columns (e) through (g) list grain abundances implied in grain models described in this work.}
\tablerefs{(1)\citet{Jenkins09}; (2)\citet{Asplund09}; (3)\citet{Preb93}}
\end{deluxetable}

(2) In our second Cloudy grain model, we explore the effects of a vacuum component. We replace some of the CMP mantle volume with a vacuum layer, thereby reducing the need for mantle elements C, N and O that are overused in model 1. Several authors have proposed that vacuum is trapped inside grains while the mantles are forming \citep[e.g.][]{Wolff94, Mathis96, Snow96}. This process leads to porous mantles. Cloudy can treat grains that are porous throughout or else it can treat grains with a vacuum layer inside; however, it cannot treat grains with porous mantles alone. To preserve the layered structure of CMPs in our grain models, we chose a vacuum layer between the silicate core and the dirty ice mantle. This choice introduces another free parameter, the fraction of the total CMP volume that is vacuum. For model 2 we chose 70\% vacuum, and we retained b/a = 1.62, as for model 1. The introduction of the vacuum layer results in a thinner mantle since the vacuum replaces much of the mantle volume. The model 2 mantle now only occupies 6\% of the CMP volume rather than 76\% as in the non-vacuum model 1. Therefore, smaller abundances of mantle elements C, N and O are required, so much so that N and O are now underused in model 2 by about a factor of two. We also increase very slightly Si/H compared to model 1. This small change results in slightly more optimal use of grain elements; however, it is otherwise insignificant. (See Table \ref{tab:appendix}, column (f).) In short, the CMPs of model 2 have the same sizes (i.e. mantle radii) as model 1 and the same silicate cores. The key difference is that much of the mantle volume in model 1 has been replaced with vacuum in model 2.

(3) In the third Cloudy grain model, to increase calculated values of \kn in the FIR while still using elements near their optimum abundances, we increased the CMP vacuum volume fraction to 80\%, and we increased b/a to 2.0 so that the CMPs are larger than in models 1 and 2. In model 3, the mantle volume is larger by a factor of about two over model 2, so the required abundances of mantle elements C, N and O are increased by this same factor. The mantle in model 3 occupies 8\% of the particle volume. We retain the same values for Si/H (hence, for Mg/H and Fe/H) as model 2.

Apart from grain model calculations such as those described above, attempts have been made to estimate FIR grain opacities from observational data, and compare to the theoretical models in the literature, including the widely used \citet{OH94} models. \citet{Shirley11} used observations of dust emission in the near-infrared (NIR, 2.2 $\micron$) and in the FIR (450 and 850 $\micron$) to estimate the opacity ratios $\kappa_{450\micron}/\kappa_{2.2\micron}$ and $\kappa_{850\micron}/\kappa_{2.2\micron}$. Including uncertainties, \citeauthor{Shirley11} find $\kappa_{450\micron}/\kappa_{2.2\micron}$ = 12--27 \e{-4} and $\kappa_{850\micron}/\kappa_{2.2\micron}$ = 2.9--5.2 \e{-4} (compare with table 5 column (f) and (g), respectively). The opacity ratios calculated for our grain models are comparable to the ranges in ratios derived by \citeauthor{Shirley11} These authors then estimate a range of grain opacities at 450 and 850 $\micron$, assuming a 2.2 $\micron$ opacity in the range 31--45 \scmpg of gas (for a gas-to-dust ratio of 100). Their opacity estimates in the two FIR wavelength bands are consistent with our calculated opacities for grain model 1. However, our grain model 3 has 2.2 $\micron$ opacity of just 10.5 \scmpg.  If this model is correct, at least at 2.2 $\micron$, then the 450 and 850 $\micron$ opacity estimates of \citeauthor{Shirley11} are about a factor three too high. Note that the opacities at these two FIR bands for \citeauthor{OH94} models fall within the range estimated by \citeauthor{Shirley11}; however, \citeauthor{OH94} models suffer from the same issue of overusing the elements as discussed for \citetalias{Preb93} model. They use similar values for C and Si abundances as those assumed by \citetalias{Preb93}. 

Model opacities in the optical range are the only ones that can be compared directly with observations. \citet{Mathis96} cites observational data indicating N(H)/E(B$-$V) = 5.8e{21} \pscm mag$^{-1}$, an average along the lines-of-sight to 45 stars. The ratio implies \kv = 68 R, where R is the usual ratio of total visual to selective extinction, and \kv is the point source extinction in the V band. Converting from point source to extended source extinctions (see footnote 2), we find \kv = 40 R. R is often taken as 3.1 for diffuse gas, and higher values of order 5 may apply to denser regions with larger grains such as in the Orion Nebula environment \citep[see][and references therein]{Abel04}. Therefore, optical observations imply that the V band (0.54 mm) extended source opacity \kv = 125--200 \scmpg. Values on the low end of this range are likely applicable to diffuse gas, values on the higher end of the range may be more appropriate for molecular clouds like the Polaris Flare with larger grains. The \citet{Draine84} extinction calculations reproduce \kv = 125 \scmpg since they were designed to be consistent with observations along lines-of-sight through diffuse gas. Our model 1, which overuses some of the grain elements, predicts \kv = 150 \scmpg, consistent with observations. However, models 2 and 3 predict \kv of only about 60 \scmpg. That is, these models that do not overuse grain elements cannot account for observed extinctions in the V band. This problem has been noted in the literature before. \citep[See][and references therein.]{Jenkins14}

A side note regarding our use of Cloudy to calculate opacities: Cloudy needs the optical data (refractive indices) in the range from about 10$^{-3} \micron$ to about 10$^4 \micron$ to calculate opacities. In the case where the data is not provided for the entire range, Cloudy extrapolates the data to calculate the opacities. In the FIR range, the scattering follows the Rayleigh limit; therefore, the opacities calculated are proportional to $\nu^2$. \citetalias{Preb93} provided the data for the grain types used in their grain model in the limited range 0.1--800 $\micron$. It is difficult to find the refractive indices in the laboratory setting for the wavelengths outside this range. Cloudy could easily extrapolate this data towards higher wavelengths. However, at lower wavelength limit, Cloudy runs into error because the extrapolated opacities are too large to be physical, and the code exits without completing the calculations.

We overcome with this problem by using the optical data from \citet{Bussoletti87} for the wavelengths smaller than 0.1 $\micron$, and modifying it such that when combined with the optical data from \citetalias{Preb93}, the opacities calculated are physical. As our ISM models do not include the photons of energy higher than ionization potential of hydrogen, 13.6 eV (or wavelengths lower than $\sim$0.1 $\micron$), the opacities calculated at lower range are not important for PDR calculations. In addition, the opacities at one wavelengths are not dependent on opacities at other wavelengths in Cloudy. Hence, this ad hoc procedure should not affect the results of the PDR calculations. For wavelengths over 800 $\micron$, we let Cloudy extrapolate the opacities. For more information on the calculation of opacities in Cloudy, please refer to \citet{c13ref} and the Cloudy documentation.

\end{document}